\documentclass[aip,twocolumn,letterpaper]{revtex4}

\usepackage{fullpage}

\usepackage{graphics}
\usepackage{epsfig}

\begin{document}
\title{Fast magnetic reconnection and the ideal evolution of a magnetic field}
\author{Allen H. Boozer}
\affiliation{Columbia University, New York, NY  10027\\ ahb17@columbia.edu}

\begin{abstract}

Regardless of how small non-ideal effects may be, phenomena associated with changes in magnetic field line connections are frequently observed to occur on an Alfv\'enic time scale.  Since it is mathematically impossible for magnetic field line connections to change when non-ideal effects are identically zero, an ideal evolution must naturally lead to states of unbounded sensitivity to non-ideal effects.  That such an evolution is natural is demonstrated by the use of Lagrangian coordinates based on the flow velocity of the magnetic field lines.  The Lagrangian representation of an evolving magnetic field  is highly constrained when neither the magnetic field strength nor the forces exerted by the magnetic field increase exponentially with time.  The development of a state of fast reconnection consistent with these constraints (1) requires a three-dimensional evolution, (2) has an exponentially increasing sensitivity to non-ideal effects, and (3) has a parallel current density, which lies in exponentially thinning but exponentially widening ribbons, with a magnitude that is limited to a slow growth. The implication is that exponential growth in sensitivity is the cause of fast magnetic reconnection when non-ideal effects are sufficiently small.  The growth of the non-ideal effect of the resistivity multiplied by the parallel current density is far too slow to be competitive.

\end{abstract}

\date{\today} 
\maketitle


\section{Introduction}

Laboratory, space, solar, and astrophysical magnetic fields are commonly embedded in highly conducting, near-ideal, plasmas.  In an ideal plasma, magnetic field lines do not change their connections or other topological properties, and the magnetic field evolves as
\begin{equation}
\frac{\partial \vec{B}}{\partial t} =\vec{\nabla}\times(\vec{u}\times\vec{B}). \label{ideal-ev}
 \end{equation} 
 with $\vec{u}$ the velocity of the magnetic field lines \cite{Newcomb}, which need not be the velocity of the plasma $\vec{v}$.  As will be shown, Equation (\ref{ideal-ev}) naturally leads to states of unbounded sensitivity to non-ideal effects.  This unbounded sensitivity eventually produces a fast change in magnetic field line connections when non-ideal effects are sufficiently small but non-zero.

Equation (\ref{ideal-ev}) for an ideal evolution can be solved exactly for both the magnetic field, $\vec{B}(\vec{x},t)$, and the current density, $\vec{j}(\vec{x},t)=\vec{\nabla}\times\vec{B}/\mu_0$ in terms of the initial magnetic field $\vec{B}_0(\vec{x}_0)$ and the flow velocity of the magnetic field lines $\vec{u}$.  The solution is based on Lagrangian coordinates, $\vec{x}_0$, which have a position vector $\vec{x}(\vec{x}_0,t)$ that is defined by
\begin{equation}
\left(\frac{\partial \vec{x}}{\partial t}\right)_{\vec{x}_0} \equiv \vec{u}(\vec{x},t) \mbox{   where   } \vec{x}(\vec{x}_0,0)=\vec{x}_0. \label{def.Lag.coord}
\end{equation}
The position vector in ordinary Cartesian coordinates is $\vec{x}=x(x_0,y_0,z_0,t)\hat{x}+y(x_0,y_0,z_0,t)\hat{y}+z(x_0,y_0,z_0,t)\hat{z}$.  

How can the introduction of an unknown velocity of the magnetic field lines $\vec{u}(\vec{x},t)$ provide important information on the evolution of magnetic fields?  The answer is that the mathematical properties of Lagrangian coordinates allow the constraints of qualitative information to be included in the analysis.
 
 The most important property of Lagrangian coordinates is exponentiation.  What is meant by this?   The Jacobian matrix of Lagragian coordinates is defined as
\begin{eqnarray}
\frac{\partial \vec{x}}{\partial \vec{x}_0} &\equiv& \left(\begin{array}{ccc}\frac{\partial x}{\partial x_0} & \frac{\partial x}{\partial y_0} & \frac{\partial x}{\partial z_0} \vspace{0.03in}  \\ \vspace{0.03in} \frac{\partial y}{\partial x_0} & \frac{\partial y}{\partial y_0} & \frac{\partial y}{\partial z_0}  \\ \frac{\partial z}{\partial x_0} & \frac{\partial z}{\partial y_0} & \frac{\partial z}{\partial z_0}\end{array}\right) \\
  &=& \tensor{U}\cdot\left(\begin{array}{ccc}\Lambda_u & 0 & 0 \\0 & \Lambda_m & 0 \\0 & 0 & \Lambda_s\end{array}\right)\cdot\tensor{V}^\dag . \label{SVD.of.Jacobian}
\end{eqnarray}
The second matrix expression for $\partial\vec{x}/\partial\vec{x}_0$ is the Singular Value Decomposition (SVD) of the first.  Any  three-by-three matrix with real coefficients can be written in this form, where $\tensor{U}$ and $\tensor{V}$ are orthogonal matrices, $\tensor{U}^\dag\cdot \tensor{U}=\tensor{1}$.  The coefficients $\Lambda_u\geq \Lambda_m \geq\Lambda_s\geq0$ of a singular value decomposition are called singular values and are positive real numbers.  For all but exceptional flow velocities $\vec{u}(\vec{x},t)$, the largest singular value $\Lambda_u$ increases exponentially with time, 
\begin{eqnarray}
\Lambda_u&=&e^{\sigma_u(\vec{x}_0,t)},   \mbox{   where   } \\
\lambda_u&\equiv&  \frac{\sigma_u}{t}
\end{eqnarray}  
is called the Lyapunov exponent in the theory of dynamical systems.  The singular value $\Lambda_s$ generally decreases exponentially  with time, $\Lambda_s=e^{-\sigma_s}$, while  $\Lambda_m$ has a much weaker evolution, sometimes a power law of time.  The time dependence of the singular values is discussed in Section \ref{sec:Lagrangian coord}.

It should be emphasized that the velocity $\vec{u}(\vec{x},t)$ must have a very special form to avoid exponentiation, not the other way around.

The ideal approximation can become inadequate for two independent reasons: (1) The non-ideal part of the electric field, typically $\eta j_{||}$, can become too large, where $\eta$ is the plasma resistivity and $j_{||}\equiv\vec{j}\cdot\vec{B}/B$.  (2) The exponentiation of the separation between neighboring magnetic field lines with distance along the lines amplifies an arbitrarily small non-ideal effect exponentially.  The trajectory of a magnetic field line can be eventually affected on the scale of the system by an arbitrarily small non-ideal effect.  

Appendix \ref{sec:B corr} shows the importance of exponentiation more directly.  While non-ideal effects on the magnetic field remain small, the direction of the magnetic field lines is perturbed by the exponentially large factor $\Lambda_u$ times the strength of the non-ideal part of the electric field.

Without constraints on $\vec{u}$, the magnetic field strength would increase as $\Lambda_u$, which increases exponentially in time, and the force exerted by the field would increase as $1/\Lambda_s$, which also increases exponentially in time.  If and only if the evolution is in three-dimensional space, constraints can be imposed to eliminate these exponential increases.   When these constraints are applied, the increase in the non-ideal effect $\eta j_{||}/B$ is proportional to the increase in $\Lambda_m^2$, which is slow.  But,  the sensitivity to non-ideal effects due to the separation of neighboring magnetic field lines increases as $\Lambda_u$.

The implication is that the effect on reconnection of the increase in the non-ideal part of the electric field becomes subdominant to the effect of exponentiation as the non-ideal effects become small.   As will be discussed, singular current densities can arise in the infinite-time limit, but even when singular currents are predicted, as in the Parker conjecture \cite{Parker:2012,Rappazzo-Parker}, their effect on reconnection is subdominant to the effect of exponentiation when non-ideal effects are sufficiently small.

A related argument on the  subdominance of current singularities to exponentiation for magnetic reconnection was given in \cite{Boozer:B-line.sep}.  There it was noted that the parallel current density $j_{||}$ required to achieve a certain number of exponentiations in the separation between neighboring magnetic field lines scales linearly in the number of exponentiations. 

Although the amplitude of $j_{||}/B$ can increase only slowly, the gradient of $j_{||}/B$ across the magnetic field lines increases exponentially in one direction and decreases exponentially in the other.  That is, $j_{||}/B$ lies in increasingly thin but wide ribbons.  Although the thinning and widening of the current sheet and fast magnetic reconnection are correlated, both are due to large scale properties of the magnetic field evolution, not a local cause-and-effect relationship.  The natural formation of current sheets in an evolving magnetic field should not be confused with the existence of Harris sheets \cite{Harris}, which are not consistent with three-dimensional plasma states \cite{Boozer:current sheets}.  Nevertheless, a Harris sheet is a standard initial condition for two-dimensional studies of magnetic reconnection.

The requirement for a three-dimensional flow for developing states of fast magnetic reconnection may at first be surprising.   Streamlines of a flow generally separate exponentially even when a time dependent velocity  $\vec{u}$ causes only a two-dimensional evolution.   However in a two-dimensional evolution, the only way to exponentially enhance reconnection is by an exponential increase in the magnetic field strength.  When the evolution is two-dimensional, say in the $x$ and $y$ coordinates, the matrix
\begin{eqnarray}
\frac{\partial \vec{x}}{\partial \vec{x}_0} &\equiv& \left(\begin{array}{ccc}\frac{\partial x}{\partial x_0} & \frac{\partial x}{\partial y_0} \vspace{0.03in}  \\ \vspace{0.03in} \frac{\partial y}{\partial x_0} & \frac{\partial y}{\partial y_0} \end{array}\right),
\end{eqnarray} 
which has only the $\Lambda_u$ and $\Lambda_s$ singular values.  In a three-dimensional evolution, the magnetic field can be proportional to the $\Lambda_m$ singular value, but that singular value is missing for a two-dimensional evolution.  The implication is that in two-dimensions, the only way to make a exponentially large change in the importance of non-ideal terms is to exponentiate the magnetic field strength.  Longcope and Strauss \cite{Longcope-Strauss}  used the two-dimensional Jacobian matrix in their 1994 discussion of the formation of current layers. 

Once magnetic field line connections are changed, large forces arise and must be relaxed.  The reason for the large forces is each of the two parts of a newly connected field line will initially have a different parallel current.  This implies a strong gradient in $j_{||}/B$ with distance along the line.  When the Debye length is small, the current density must be divergence free \cite{Boozer:NF3D}, and $\vec{\nabla}\cdot\vec{j}=0$ can be written as
\begin{eqnarray}
\vec{B}\cdot\vec{\nabla} \frac{j_{||}}{B} &=&\vec{B}\cdot\vec{\nabla}\times\frac{\vec{f}_L}{B^2},  \mbox{   where   } \\
\vec{f}_L &\equiv& \vec{j}\times\vec{B}.
\end{eqnarray}
The Lorentz force, $\vec{f}_L$ is the force of the magnetic field on the plasma.  The sudden change in the force on the plasma is balanced by the plasma viscosity, $\vec{f}_L=\vec{\nabla}\cdot\tensor{P}$ or by the plasma inertia $\vec{f}_L=\rho(\partial\vec{v}/\partial t +\vec{v}\cdot\vec{\nabla}\vec{v})$.  The inertial term implies a relaxation by Alfv\'en waves.  An Alfv\'enic limit on the speed of reconnection effects is not surprising---that is the fastest speed at which magnetic fields can transmit information either along or across magnetic field lines in the standard MHD approximation.  Nevertheless, the time required for an Alfv\'enic relaxation is subtle because Alfv\'en waves propagating along exponentially separating field lines have enhanced damping \cite{Heyvaerts-Priest:1983,Similon:1989}. 

Magnetic field lines are commonly observed to change their connections on a time scale consistent with the Alfv\'en speed $V_A$.  A review of the observations and theory of reconnection proceeding at $0.1V_A$ has been given by Cassak et al \cite{Cassak:2017}.   Reconnection at a rate closely associated with Alfv\'enic rather than non-ideal effects is called fast magnetic reconnection.  Although Equation (\ref{ideal-ev}) for an ideal evolution is inconsistent with magnetic field lines changing their topology, the equation predicts an exponentially increasing sensitivity to non-ideal effects,  Appendix \ref{sec:B corr}, which leads to fast magnetic reconnection. 

In the limit as non-ideal effects are very small, a fast reconnection event will occur after an adequate time, called a trigger time, for the ideal evolution to produce a sufficiently large $\Lambda_u$. 

Assuming the resistivity $\eta$ is the most important non-ideal term, the degree of exponentiation required for a reconnection trigger is $\Lambda_u \sim \Im$.   The dimensionless coefficient that measures the closeness to ideality,  a Gothic  ``I,"   is
\begin{eqnarray}
\Im &\equiv& \frac{\tau_\eta}{\tau_{ev}},  \\
\tau_\eta &=& \frac{\mu_0}{\eta} a^2,
\end{eqnarray}
is the time required for resistive diffusion across of the magnetic field lines on the scale $a$ of the region that undergoes reconnection, and $\tau_{ev} \equiv 1/|\vec{\nabla} \vec{u}|$  is the characteristic time scale for the magnetic evolution.   When $\Im\lesssim 1$, resistive diffusion is so rapid that the difference between two and three-dimensional reconnection is of limited importance.  As $\Im$ becomes larger, the physics of reconnection becomes ever more sensitive to any breaking of the continuous symmetry that is assumed in two-dimensional models.   The derivations of this paper are in the limit as $\Im\rightarrow \infty$, and important work remains to be done in how the validity of a two-dimensional models breaks down as $\Im$ increases.

Remarkably, the importance of exponentiation has largely escaped notice in the literature on magnetic reconnection.  For example, major recent reviews \cite{Zweibel:review,Loureiro:2016} focused on two-dimensional plasmoid models of magnetic reconnection, which are descendants  of the 1957 models of Sweet and Parker \cite{Sweet:1958,Parker:1957}.   An explanation for the speed of magnetic reconnection in a two dimensional plasmoid model based on a Harris current sheet warranted a 2017 Physical Review Letter \cite{Liu:2017}.  

When non-ideal effects are small, the mathematical properties of an evolving magnetic field are fundamentally different between two and three dimensions.   These differences bring the relevance of two-dimensional plasmoid models to three-dimensional problems into question.   

Plasmoids are analogous to magnetic islands in topologically toroidal plasmas.  Islands arise when perfect toroidal magnetic surfaces are perturbed.  But, islands are highly localized in toroidal plasmas; they only split magnetic surfaces that are rational surfaces, surfaces on which magnetic field lines close on themselves.  Otherwise Alfv\'en waves spread the effect of a perturbation over the volume of space covered by a single field line, and an island is not formed.  How this spreading is consistent with plasmoid formation in three-dimensional space remains to be explained.

Delta function current densities are mathematically required when a rational magnetic surface in an ideal, steady-state, toroidal plasma is resonantly perturbed \cite{Boozer-Pomphrey}. Nevertheless, as shown by Hahm and Kulsrud \cite{Hahm-Kulsrud}, the current density in the vicinity of the rational surface increases only linearly in time after a resonant perturbation is applied.  This arises from the time required for a shear-Alfv\'en wave propagating along the magnetic field to cover a near-rational surface and adequately sample its topological properties.  The Hahm and Kulsrud  time is proportional to $1/(N-\iota M)$, where the magnetic field lines on the rational surface close on themselves after $M$ toroidal and $N$ poloidal transits.  The rotational transform $\iota$ or twist of the magnetic field lines at the rational surface is $N/M$.  The applicability of plasmoid models to reconnection in three-dimensional plasmas cannot be understood unless the analogue of the Hahm-Kulsrud time for the bounding surface of a plasmoid is obtained.  Appendix \ref{RMHD} gives another example of a linear increase in the current density as a magnetic field is evolved ideally but slowly compared to the Alfv\'en speed.

What will be studied in this paper is the nature of magnetic reconnection in the limit as non-ideal effects go to zero, the limit as $1/\Im$ goes to zero.  Turbulence produces three-dimensionality \cite{Turbulent reconnection} and can give the exponential sensitivity required for fast reconnection as non-ideal effects become arbitrarily small, but turbulence is not required.  Indeed, three-dimensionality on a large spatial scale causes the reconnection to occur on a large scale.  Fast magnetic reconnection works much like stirring, Section \ref{sec:Lag.appl.}.  Large-scale stirring is a more effective way to mix a can of paint than small-scale stirring though both can cause mixing.

The sensitivity of the ideal constraint on magnetic evolution near places where current density becomes large has been discussed by a number of authors, including Low \cite{Low:2015} and Dewar et al \cite{Dewar:2017}.  Here it will be shown that in a highly ideal evolution the sensitivity is given by the magnitude of $\Lambda_u$ and not directly by the current density.   

A motivation for writing this paper  was the development of an understanding of the rapid loss, $\lesssim 1~$ms, of magnetic surfaces that is commonly observed during the thermal quench phase of a tokamak disruptions.  For example, in JET \cite{de Vries:2016} the growth of resonant magnetic perturbations was observed to occur over 100's of ms before the current profile suddenly broadened reducing the internal inductance $\ell_i$ by approximately a factor of two.  The relation between these observations and fast magnetic reconnection is discussed in \cite{Boozer:pivotal,Boozer:prevalence}.  In JET, the growth of islands over 100's of ms is consistent with the resistive opening of islands, but the sudden and large change in the current profile is not.

Section \ref{sec:conditions} derives the conditions for an ideal evolution.  Section \ref{sec:Lagrangian coord} derives the properties of Lagrangian coordinates required to determine the properties of ideally evolving magnetic fields.  Section \ref{sec:ev of B} obtains the magnetic field evolution in Lagrangian coordinates.   Section \ref{sec:j-f ev.} gives expressions for the current density and the Lorentz, $\vec{j}\times\vec{B}$, force. Section \ref{sec:B-lines} obtains the separation of the magnetic field lines.   Section \ref{sec:Lag.appl.} discusses applications of Lagrangian coordinates in other areas of classical physics.  Section \ref{sec:Discussion} is a discussion, which focuses on the distinction between standard two-dimensional models of reconnection and the theory developed here.   There are two appendices.  Appendix \ref{sec:B corr} derives the exponentially increasing departure of a magnetic field from its ideal form under the assumption that that departure is small.  Appendix \ref{RMHD} derives the current density in the early stage of an ideal evolution when the system is driven by a slow flow in perfectly  conducting wall.



\section{Conditions for an ideal evolution \label{sec:conditions} }


\subsection{Ohm's law}

Equation (\ref{ideal-ev}) for the ideal evolution of a magnetic field holds in regions in which a magnetic field has nulls, $\vec{B}=0$ points, when the plasma moving with a velocity $\vec{v}$ obeys an Ohm's law $\vec{E}+\vec{v}\times\vec{B}=0$.  Then, the field line velocity is the plasma velocity $\vec{u}\equiv\vec{v}$.   

The ideal evolution equation has a more general validity in regions of space in which there are no nulls of $\vec{B}$.  The magnetic field line velocity $\vec{u}$ can be obtained from the generalized Ohm's law
\begin{equation}
\vec{E}+\vec{v}\times\vec{B} = \eta_{||}\vec{j}_{||} + \eta_\bot \vec{j}_\bot +\frac{\vec{j}\times\vec{B}}{en} - \frac{\vec{\nabla}\cdot \tensor{p}_e}{en}+\frac{m_e}{e^2n} \frac{\partial \vec{j}}{\partial t}, 
\end{equation}
which implies the velocity of the fluid mass $\vec{v}$ perpendicular to magnetic field lines that move with a velocity $\vec{u}_\bot$  satisfies
\begin{eqnarray}
\vec{v}_\bot &=&\vec{u}_\bot-\eta_\bot \frac{\vec{j}\times\vec{B}}{B^2} -\frac{\vec{j}_\bot}{en}-\frac{\vec{B}\times\vec{\nabla}\cdot \tensor{p}_e}{enB^2}\nonumber\\
&&+\vec{B}\times\left(\frac{m_e}{e^2n} \frac{\partial \vec{j}}{\partial t} \right)-\frac{\vec{B}\times\vec{\nabla}\Phi}{B^2}, \label{fluid velocity}
\end{eqnarray}
where $\Phi$ is an arbitrary single-valued function of position.  When the inertial force $\rho d\vec{v}/dt$ is negligible,, the plasma velocity $\vec{v}$ does not directly enter the evolution of the magnetic field.  Nevertheless, the plasma flow can have a peculiar form relative to that of the field because of what are called Hall terms in the generalized Ohm's law.   

The only physical effects that break the ideal evolution of the magnetic field are on the right hand side of the equation
\begin{eqnarray}
 &&\vec{E}+\vec{u}\times\vec{B} +\vec{\nabla}\Phi = \mathcal{E}_{ni} \vec{\nabla}\ell, \mbox{   where   } \label{gnl.Ohms} \\
&& \mathcal{E}_{ni} \equiv \frac{\vec{B}}{B}\cdot\left( \eta_{||}\vec{j}+\frac{m_e}{e^2n} \frac{\partial \vec{j}}{\partial t} - \frac{\vec{\nabla}\cdot \tensor{p}_e }{en} - \vec{\nabla}\Phi \right). \hspace{0.2in}
 \end{eqnarray}
 The potential $\Phi$ can be chosen to minimize the field-line constant $\mathcal{E}_{ni}$ with
 \begin{equation}
 \mathcal{E}_{ni} = \frac{\int \vec{E}\cdot d\vec{\ell}}{\int d\ell}.
 \end{equation}
The integration limits on the two integrals are the same with $d\vec{\ell}$ the differential vector distance along a magnetic field line and $d\ell$ the differential scalar distance.  The integration limits can be plus and minus $L\rightarrow\infty$ or from one perfectly conducting boundary to another.  The ideality of the evolution can be measured by $\Im=uB/ \mathcal{E}_{ni}$.


\subsection{The Clebsch representation \label{sec:Clebsch rep} }

A magnetic field can always be given in the Clebsch representation, 
\begin{eqnarray}
\vec{B}(\vec{x},t)&=&\vec{\nabla}\alpha(\vec{x},t)\times \vec{\nabla}\beta(\vec{x},t),  \mbox{  so   }\\
\left(\frac{\partial\vec{B}}{\partial t}\right)_{\vec{x}} &=& \vec{\nabla}\times\left(\frac{\partial\alpha}{\partial t}\vec{\nabla}\beta -\frac{\partial\beta}{\partial t}\vec{\nabla}\alpha +\vec{\nabla}\alpha\frac{\partial\beta}{\partial t}\right). \hspace{0.2in}
\end{eqnarray}
A particular magnetic field line is specified by constant values of $\alpha$ and $\beta$.  Stern \cite{Stern:1970} has reviewed the history of the Clebsch representation.

The Clebsch potentials $\alpha$ and $\beta$ can be used with the distance along the magnetic field lines $\ell$ to form a spatial coordinate system, which means the three Cartesian coordinates $(x,y,z)$ with  $\vec{x}=x\hat{x}+y\hat{y} +z\hat{z}$ are given as functions of $(\alpha,\beta,\ell)$.  Clebsch coordinates are a subtle when going from Cartesian space to Clebsch space, but not the other way around.  When $\alpha$ and $\beta$ are held fixed, $\vec{x}(\alpha,\beta,\ell)$ gives every point in Cartesian coordinates that a particular magnetic field line can reach as $\ell$ is varied.  It is this property that makes the Clebsch variables $(\alpha,\beta,\ell)$ useful for determining the deviation of a magnetic field from its ideal form due to a non-zero $\mathcal{E}_{ni}$, Appendix \ref{sec:B corr}.    

Faraday's law, $\partial\vec{B}/\partial t=-\vec{\nabla}\times\vec{E}$, can be written using Equation \ref{gnl.Ohms} for the electric field as $\partial\vec{B}/\partial t=\vec{\nabla}\times(\vec{u}\times\vec{B}-\mathcal{E}_{ni}\vec{B}/B)$.  Since $\vec{u}\times\vec{B}=(\vec{u}\cdot\vec{\nabla}\beta)\vec{\nabla}\alpha - (\vec{u}\cdot\vec{\nabla}\alpha)\vec{\nabla}\beta$,
\begin{eqnarray}
&& \frac{\partial\alpha}{\partial t}\vec{\nabla}\beta -\frac{\partial\beta}{\partial t}\vec{\nabla}\alpha=(\vec{u}\cdot\vec{\nabla}\beta)\vec{\nabla}\alpha - (\vec{u}\cdot\vec{\nabla}\alpha)\vec{\nabla}\beta \hspace{0.2in} \nonumber \\
&& \hspace{1.2in} -\mathcal{E}_{ni}(\alpha,\beta,t)\vec{\nabla}\ell +\vec{\nabla}g. 
\end{eqnarray}
Since $\vec{B}\cdot\vec{\nabla}\ell=B$, the component of this equation along the magnetic field is
\begin{eqnarray}
&&\vec{B}\cdot\vec{\nabla}g=\mathcal{E}_{ni} B,  \mbox{   or   } \left(\frac{\partial g}{\partial\ell}\right)_{\alpha\beta}= \mathcal{E}_{ni},  \mbox{   so  }\\
&& g =\mathcal{E}_{ni}(\alpha,\beta,t)\ell + g_a(\alpha,\beta,t).
\end{eqnarray}
The evolution equations for $\alpha$ and $\beta$ are
\begin{eqnarray}
\left(\frac{\partial\alpha}{\partial t}\right)_{\vec{x}} &=&-\vec{u}\cdot\vec{\nabla}\alpha + \frac{\partial g}{\partial \beta} \\
\left(\frac{\partial\beta}{\partial t}\right)_{\vec{x}} &=&-\vec{u}\cdot\vec{\nabla}\beta- \frac{\partial g}{\partial \alpha}.
\end{eqnarray}
A coordinate transformation to Lagrangian coordinates gives the equations
\begin{eqnarray}
\left(\frac{\partial\alpha}{\partial t}\right)_{\vec{x}_0} &=& \frac{\partial g}{\partial \beta} \\
\left(\frac{\partial\beta}{\partial t}\right)_{\vec{x}_0} &=&- \frac{\partial g}{\partial \alpha}.
\end{eqnarray}
The arbitrary function $g_a(\alpha,\beta,t)$ just couples $\alpha$ and $\beta$ but that does not change the identification of a magnetic field line with fixed $\alpha$ and $\beta$ as the line is carried by the flow; $g_a$ can be chosen arbitrarily.  The term $\mathcal{E}_{ni}\ell$ in $g$ breaks the magnetic field lines from the flow and can change their topology.   When $\mathcal{E}_{ni}=0$, the Clebsch coordinates are functions of the Lagrangian coordinates alone $\alpha(\vec{x}_0)$ and $\beta(\vec{x}_0)$.  

A non-trivial example of $\alpha$ and $\beta$ either satisfying the required equations for an ideal evolution, or not, is given by a toroidal plasma with magnetic surfaces.  The toroidal magnetic flux enclosed by a surface, $\psi_t$, the poloidal angle, $\theta$, and the toroidal angle, $\varphi$, can be chosen so $\alpha = \psi_t$ and $2\pi\beta=\theta-\iota\varphi$, where  $\iota(\psi_t,t)$ is called the rotational transform.   When the toroidal field is strong $\ell= R_0\varphi$ and one can assume $(\partial\ell/\partial t)_{\vec{x}}=0$.  The non-ideal part of $g=R_0\mathcal{E}_{ni}(\psi_t,t)\varphi$.  The evolution equations for $\alpha$ and $\beta$, then imply $(\partial\psi_t/\partial t)_{\vec{x}}=-\vec{u}\cdot\vec{\nabla}\psi_t$ and $(\partial\theta/\partial t)_{\vec{x}}=-\vec{u}\cdot\vec{\nabla}\theta$.  When $(\partial\iota/\partial t)_{\psi_t}\neq0$, the evolution equation for $\beta$ implies $\varphi(\partial\iota/\partial t)_{\psi_t} = 2\pi\partial g/\partial\psi_t$, or $(\partial\iota/\partial t)_{\psi_t}=2\pi R_0(\partial\mathcal{E}_{ni}/\partial \psi_t)$.  The loop voltage is $V_{\ell} =2\pi R_0\mathcal{E}_{ni}$, so $(\partial\iota/\partial t)_{\psi_t}=\partial V_\ell/\partial\psi_t$, which is a well-known result in the physics of toroidal plasmas \cite{Boozer:NF3D}.



\section{ Lagrangian coordinates \label{sec:Lagrangian coord} }

Lagrangian coordinates $\vec{x}_0$, which were defined in Equation (\ref{def.Lag.coord}), have a number of general properties that are required to understand their implications for magnetic evolution.  These properties are derived in this section.

\subsection{The Jacobian matrix} 

Equation (\ref{SVD.of.Jacobian}) for the SVD decomposition of the three-by-three matrix $\partial\vec{x}/\partial\vec{x}_0$ is equivalent to
\begin{equation}
\frac{\partial \vec{x}}{\partial \vec{x}_0} = \hat{U}\Lambda_u\hat{u}+\hat{M}\Lambda_m\hat{m}+\hat{S}\Lambda_s\hat{s}. \label{Jacobian rep}
\end{equation}
The left eigenvectors, which are defined by the orthogonal matrix $\tensor{U}$ obey
\begin{equation}
\hat{U}\cdot\hat{U}=1\mbox{  etc. and  } \hat{U}=\hat{M}\times\hat{S}.
\end{equation}
The right eigenvectors, which are defined by the orthogonal matrix $\tensor{V}$, obey analogous relations
\begin{equation} 
\hat{u}\cdot\hat{u} = 1 \mbox{   etc. and  }\hat{u}=\hat{m}\times\hat{s}.
\end{equation}
The Jacobian of Lagrangian coordinates, which is the determinant of the matrix $\partial\vec{x}/\partial\vec{x}_0$ obeys
\begin{equation}
J=\Lambda_u\Lambda_m\Lambda_s.
\end{equation}

The time derivative of the Jacobian can be found by considering the evolution of an arbitrary volume defined in fixed Lagrangian coordinates.  Then,
\begin{eqnarray}
\frac{\partial \mbox{Vol} }{\partial t} &=&\oint \vec{u}\cdot d \vec{a} = \int \vec{\nabla}\cdot\vec{u} d^3x=\int J \vec{\nabla}\cdot\vec{u} d^3x_0\nonumber\\
&=& \int \left( \frac{\partial J}{\partial t}\right)_{\vec{x}_0} d^3x_0, \mbox{   so   } \\
\left(\frac{\partial J}{ \partial t}\right)_{\vec{x}_0} &=& J \vec{\nabla}\cdot \vec{u}. \label{dJ/dt}
\end{eqnarray}

The component of the magnetic field line velocity $\vec{u}$ in the direction along $\vec{B}$ can be defined freely.  Letting $\hat{b}=\vec{B}/B$,  three attractive choices are:  (1) $\hat{b}\cdot\vec{u}=0$, which is the simplest and the choice used in this paper, (2) $\vec{u}\cdot\vec{\nabla}\ell=0$, which makes $u_{||}=-\vec{u}_\bot\cdot\vec{\nabla}\ell$, and (3)  $\hat{b}\cdot\vec{\nabla}(\hat{b}\cdot\vec{u}) = \vec{u}\cdot (\hat{b}\cdot\vec{\nabla}\hat{b})$, which makes $JB=B_0$,  The requirement for the $JB=B_0$ condition was incorrectly stated  in \cite{Boozer:prevalence} as $\hat{b}\cdot\vec{u}=0$, which is only true where the curvature of the magnetic field lines $\hat{b}\cdot\vec{\nabla}\hat{b}$ vanishes.  


\subsection{Evolution of the Jacobian matrix}

To clarify the meaning of complicated dot products, Greek letters $\alpha, \beta,\gamma, \cdots$ will be used to denote ordinary Cartesian coordinates $\vec{x}$ and Latin letters $i,j,k,\cdots$ for Lagrangian coordinates $\vec{x}_0$.  Equation (\ref{def.Lag.coord}) for the definition of Lagrangian coordinates implies
\begin{eqnarray}
\frac{\partial}{\partial t}\left(\frac{\partial x^\alpha(\vec{x}_0,t)}{\partial x_0^i}\right) &=& \sum_\beta \frac{\partial x^\beta}{\partial x_0^i}\frac{\partial u^\alpha}{\partial x^\beta} \label{first d tensor(J)/dt} \\
&=& \sum_\beta G^\alpha_\beta \frac{\partial x^\beta}{\partial x_0^i} ,  \mbox{   where   }  \label{d(tensor(J)/dt}\\
G^\alpha_\beta &\equiv& \frac{\partial u^\alpha}{\partial x^\beta}.
\end{eqnarray}
In matrix notation
\begin{eqnarray}
\frac{\partial}{\partial t}\frac{\partial \vec{x}}{\partial x_0^i} = \tensor{G}\cdot\frac{\partial\vec{x}}{\partial x_0^i} \label{Ev.Jacobian.Matrix}
\end{eqnarray}


\subsection{The metric tensor}

The exponential increase in $\Lambda_u$ is made plausible by the expression for the evolution of the distance $\vec{\delta}$ between neighboring (infinitesimally separated) streamlines.  As $|\vec{\delta}|\rightarrow0$, 
\begin{eqnarray}
\frac{d(\vec{x}+\vec{\delta})}{dt} &=& \vec{u}(\vec{x}+\vec{\delta},t) \mbox{   so   }\\
\frac{d\vec{\delta}}{dt} &=& \vec{\delta}\cdot \vec{\nabla}\vec{u}(\vec{x},t); \label{sep.ev}  \\
\vec{\delta}&=&\frac{\partial \vec{x}}{\partial\vec{x}_0}\cdot \delta\vec{x}_0.
\end{eqnarray}
Equation (\ref{sep.ev}) is a linear equation for $\vec{\delta}$, which generally has an exponentially increasing solution.

The distance between neighboring stream lines is
\begin{eqnarray}
\vec{\delta}\cdot \vec{\delta} &=& \delta\vec{x}^\dag_0\cdot \left(\frac{\partial \vec{x}}{\partial\vec{x}_0}\right)^\dag \cdot \left(\frac{\partial \vec{x}}{\partial\vec{x}_0}\right) \cdot \delta\vec{x}_0. \label{eq:distance}
\end{eqnarray}

The symmetric tensor that appears in Equation(\ref{eq:distance}) is the metric tensor,
\begin{eqnarray}
\tensor{g} &\equiv&  \left(\frac{\partial \vec{x}}{\partial\vec{x}_0}\right)^\dag \cdot \left(\frac{\partial \vec{x}}{\partial\vec{x}_0}\right) \label{metric tensor def}\\
\nonumber\\
&=& \Lambda_u^2 \hat{u} \hat{u}+\Lambda_m^2 \hat{m}\hat{m}  + \Lambda_s^2 \hat{s} \hat{s}. \label{metric tensor rep}
\end{eqnarray}


\subsection{Evolution of the metric tensor}

The time derivative of the metric tensor can be written as
\begin{equation}
\left(\frac{\partial \tensor{g}}{\partial t}\right)_{\vec{x}_0} = \tensor{\mathcal{G}}.
\end{equation}

Using Equation (\ref{metric tensor def}) for the metric tensor and Equation (\ref{Ev.Jacobian.Matrix}) for the time derivative of the Jacobian matrix
\begin{eqnarray}
\frac{\partial\tensor{g}}{\partial t} &=& \left(\frac{\partial}{\partial t} \frac{ \partial\vec{x} }{\partial\vec{x}_0}\right)^\dag \cdot \left(\frac{\partial\vec{x}}{\partial\vec{x}_0}\right) + \left(\frac{ \partial\vec{x} }{\partial\vec{x}_0}\right)^\dag \cdot \left(\frac{\partial}{\partial t}  \frac{\partial\vec{x}}{\partial\vec{x}_0}\right) \nonumber\\
&=& \left(\frac{ \partial\vec{x} }{\partial\vec{x}_0}\right)^\dag \cdot (\tensor{G}+\tensor{G}^\dag)\cdot \left(\frac{\partial\vec{x}}{\partial\vec{x}_0}\right) \mbox{   so   } \\  
\nonumber\\
 \tensor{\mathcal{G}}&\equiv& \left(\frac{ \partial\vec{x} }{\partial\vec{x}_0}\right)^\dag \cdot (\tensor{G}+\tensor{G}^\dag)\cdot \left(\frac{\partial\vec{x}}{\partial\vec{x}_0}\right) 
\end{eqnarray}


\subsection{Evolution of $\Lambda_{u,m,s}^2$, $\hat{u}$, $\hat{m}$, and $\hat{s}$ \label{sec:right unit vec relax} }

The evolution of the basis vectors of Lagrangian coordinates was derived in 1987 by Goldhirsch,  Sulem, and Orszag  \cite{Goldhirsch:1987}.  Contributions were also made by Tang and Boozer \cite{Tang-Boozer:1996,Tang-Boozer:1997}, by Thiffeault \cite{Thiffeault:2002}, and by Theffeault and Boozer \cite{Thiffeault-Boozer:2003}.  These techniques were applied to the evolution of the magnetic field in a dynamo in 2000 by Tang and Boozer \cite{Tang-Boozer:2000}  and in 2003 by Thiffeault and Boozer \cite{Thiffeault-Boozer:2003}.

To determine the time dependence of the $\hat{u}$ basis vector holding Lagrangian coordinates constant, calculate
\begin{eqnarray}
\frac{\partial \tensor{g}\cdot \hat{u}}{\partial t}&=& \tensor{\mathcal{G}}\cdot\hat{u} + \tensor{g}\cdot\frac{\partial \hat{u}}{\partial t}. \label{d(g.e)/dt}
\end{eqnarray}
Using Equation (\ref{metric tensor rep}) for $\tensor{g}$ expanded in its basis vectors,
\begin{eqnarray}
\frac{\partial \tensor{g}\cdot \hat{u}}{\partial t}&=& \frac{\partial \Lambda_u^2}{\partial t} \hat{u}  + \Lambda_u^2 \frac{\partial \hat{u}}{\partial t}.
\end{eqnarray}
Since $\hat{u}\cdot\hat{u}=1$, $\hat{u}\cdot \partial \hat{u}/\partial t =0$, the dot products of the unit vectors $\hat{u}$, $\hat{m}$, and $\hat{s}$ with Equation (\ref{d(g.e)/dt}) imply
\begin{eqnarray}
\left(\frac{\partial \Lambda_u^2}{\partial t}\right)_{\vec{x}_0} &=& \hat{u}\cdot\tensor{\mathcal{G}}\cdot\hat{u} \nonumber\\
&=& \Lambda_u^2 \hat{U}\cdot (\tensor{G}+\tensor{G}^\dag)\cdot\hat{U}; \label{Lambda_u ev}\\
\hat{m}\cdot \left(\frac{\partial \hat{u}}{\partial t} \right)_{\vec{x}_0} &=& \frac{1}{\Lambda_u^2}\left(\hat{m}\cdot\tensor{\mathcal{G}}\cdot\hat{u} + \Lambda_m^2 \hat{m}\cdot \frac{\partial \hat{u}}{\partial t}\right) \label{m.du/dt}\nonumber \\
&=& \frac{\hat{m}\cdot\tensor{\mathcal{G}}\cdot\hat{u} }{\Lambda_u^2-\Lambda_m^2} \nonumber \\
&=& \frac{\hat{M}\cdot (\tensor{G}+\tensor{G}^\dag)\cdot\hat{U} }{\frac{\Lambda_u}{\Lambda_m}- \frac{\Lambda_m}{\Lambda_u} };\\
\hat{s}\cdot \left(\frac{\partial \hat{u}}{\partial t} \right)_{\vec{x}_0} &=& \frac{\hat{S}\cdot (\tensor{G}+\tensor{G}^\dag)\cdot\hat{U} }{\frac{\Lambda_u}{\Lambda_s}- \frac{\Lambda_s}{\Lambda_u} }
\end{eqnarray}

When the dot products of $\tensor{G}+\tensor{G}^\dag$ are not increasing as rapidly as $\Lambda_u$ or $1/\Lambda_s$, the unit vector $\hat{u}$ has no further evolution once $\Lambda_u>>\Lambda_m>>\Lambda_s$.  An exponential increase in the $\tensor{G}+\tensor{G}^\dag$ dot products requires an exponential increase in $|\vec{\nabla}\vec{u}|^2 \equiv \sum_{\alpha\beta}(\partial u^\alpha/\partial x^\beta)^2$.

The evolution of $\hat{m}$ and $\hat{s}$ can be found by appropriate changes in the letters that appear in the formula for the evolution of $\hat{u}$.  For example,
\begin{equation}
\hat{s}\cdot \left( \frac{\partial\hat{m}}{\partial t} \right)_{\vec{x}_0}  =\frac{\hat{S}\cdot (\tensor{G}+\tensor{G}^\dag)\cdot\hat{M} }{\frac{\Lambda_m}{\Lambda_s}- \frac{\Lambda_s}{\Lambda_m}}
\end{equation} 
The orthogonality of the unit vectors, $\hat{u}\cdot\hat{m}=0$ implies $\hat{u}\cdot \partial \hat{m}/\partial t=-\hat{m}\cdot \partial \hat{u}/\partial t$

The evolution of the large singular value, $\Lambda_u$ is determined by Equation (\ref{Lambda_u ev}).   The properties of this evolution can be understood by noting that $\tensor{G}+\tensor{G}^\dag$ is a Hermitian matrix so it can be diagonalized with real eigenvalues,
\begin{eqnarray}
\frac{\tensor{G}+\tensor{G}^\dag}{2} &=& \tensor{Z}\cdot  \left(\begin{array}{ccc}\nu_+ & 0 & 0 \\0 & \nu_0 & 0 \\0 & 0 & \nu_-\end{array}\right) \cdot\tensor{Z}^\dag \\
&=&\hat{Z}_+ \nu_+ \hat{Z}_+ + \hat{Z}_0 \nu_0 \hat{Z}_0 + \hat{Z}_- \nu_- \hat{Z}_-
\end{eqnarray}
since $\tensor{Z}$ is an orthogonal matrix.  The three eigenvalues are are rates, with units of one over time, and ordered so $\nu_+\geq\nu_0\geq\nu_-$.  For a divergence-free flow they must sum to zero, $\nu_+ + \nu_0+\nu_-=0$.

The unit vector $\hat{U}$ rotates, Section \ref{U et al ev}, in such a way to come into alignment with $\hat{Z}_+$; at any instant
\begin{eqnarray}\hat{U}&=&\cos\theta \hat{Z}_+  + \sin\theta \cos\varphi \hat{Z}_0 + \sin\theta \sin\varphi \hat{Z}_- ; \hspace{0.2in} 
\end{eqnarray}
\begin{eqnarray}
\left(\frac{\partial \ln\Lambda_u}{\partial t}\right)_{\vec{x}_0} &=& \nu_+ \cos^2\theta + \nu_0 \sin^2\theta\cos^2\varphi \nonumber\\
&& \hspace{0.2in}+ \nu_- \sin^2\theta\sin^2\varphi \nonumber\\
&=& \nu_{ef}(\vec{x}_0,t),
\end{eqnarray}
where the effective rate of growth of $\ln\Lambda_u$ satisfies $\nu_{ef}\leq \nu_+$.   The infinite time Lyapunov exponent is defined as 
\begin{equation}
\lambda_{\infty}(\vec{x}_0) \equiv \lim_{T\rightarrow\infty} \frac{\int_0^T \nu_{ef} dt}{T}.
\end{equation}
Let $\nu_{ef}= \lambda_{\infty} +\tilde{\nu}_{ef}$, then when $\int_0^\infty \tilde{\nu}_{ef}dt = \ln{A}$ the largest singular value has the form
\begin{equation}
\Lambda_u = A(\vec{x}_0) e^{\lambda_\infty t}. \label{Lamda_u dependence}
\end{equation}
This form is equivalent to Equation (1.11) of Goldhirsch et al  \cite{Goldhirsch:1987} though other forms are possible in which the amplitude $A$ has a sufficiently weak time dependences that $(\ln A)/t$ goes to zero as $t$ goes to infinity.  Reference \cite{Tang-Boozer:1996} has a detailed discussion of these issues for two-dimensional divergence-free flows.
 

\subsection{Evolution of $\hat{U}$, $\hat{M}$, and $\hat{S}$ \label{U et al ev} }

Equation (\ref{Ev.Jacobian.Matrix})  for the time derivative of the Jacobian matrix can be used with Equation (\ref{Jacobian rep}), which gives the representation of the Jacobian matrix, to obtain the time derivatives of the $\hat{U}$, $\hat{M}$, and $\hat{S}$ unit vectors.  Taking the time derivatives with fixed Lagrangian coordinates,
\begin{eqnarray}
\frac{\partial}{\partial t} \frac{\partial \vec{x}}{\partial \vec{x}_0} &=& \frac{\partial \Lambda_u \hat{U}}{\partial t} \hat{u} + \Lambda_u \hat{U}\frac{\partial\hat{u}}{\partial t}  \nonumber \\
&& + \frac{\partial\Lambda_m \hat{M}}{\partial t} \hat{m} + \Lambda_m \hat{M} \frac{\partial\hat{m} }{\partial t} \nonumber \\
&& + \frac{\partial\Lambda_s \hat{S}}{\partial t} \hat{s} + \Lambda_s \hat{S} \frac{\partial\hat{s} }{\partial t}.
\end{eqnarray}
The time derivative of the Jacobian matrix can also be written as
\begin{eqnarray}
\frac{\partial}{\partial t} \frac{\partial \vec{x}}{\partial \vec{x}_0} &=&\tensor{G}\cdot\frac{\partial\vec{x}}{\partial \vec{x}_0} \\
&=& (\tensor{G}\cdot\hat{U}) \Lambda_u \hat{u} + (\tensor{G}\cdot\hat{M})\Lambda_m \hat{m} \nonumber \\
&&+(\tensor{G}\cdot\hat{S}) \Lambda_s \hat{s}.
\end{eqnarray}
Dotting on the left with $\hat{M}$ and on the right with $\hat{u}$ gives
\begin{equation}
\Lambda_u \hat{M}\cdot \left( \frac{\partial\hat{U}}{\partial t} \right)_{\vec{x}_0} +\Lambda_m \hat{u}\cdot \left( \frac{\partial \hat{m}}{\partial t}\right)_{\vec{x}_0} = \Lambda_u \hat{M}\cdot\tensor{G}\cdot\hat{U}.
\end{equation}
Equation (\ref{m.du/dt}) for $\hat{m}\cdot\partial \hat{u}/\partial t=-\hat{u}\cdot\partial \hat{m}/\partial t$, then implies
\begin{equation}
\hat{M}\cdot \left(\frac{\partial \hat{U}}{\partial t} \right)_{\vec{x}_0} = \frac{ \hat{M}\cdot\tensor{G}\cdot\hat{U}+\frac{\Lambda_m^2}{\Lambda_u^2} \hat{M}\cdot\tensor{G}^\dag \cdot \hat{U} }{1-\frac{\Lambda_m^2}{\Lambda_u^2} }.
\end{equation}
When $\Lambda_u>>\Lambda_m$, $\hat{M}\cdot (\partial \vec{U}/\partial t) = \hat{M}\cdot\tensor{G}\cdot\hat{U}$, which implies $\hat{U}$ has a continual rotational evolution.


\section{Evolution of the magnetic field \label{sec:ev of B} } 

Equation (\ref{ideal-ev}) for the ideal evolution of a magnetic field can be solved solved using Lagrangian coordinates, Equation (\ref{def.Lag.coord}), 
\begin{equation}
\vec{B}(\vec{x},t) = \frac{1}{J}\frac{\partial \vec{x}}{\partial\vec{x}_0}\cdot\vec{B}_0(\vec{x}_0). \label{B exp}
\end{equation}
$J$ is the Jacobian of Lagragian coordinates, which is the determinant of the matrix $\partial\vec{x}/\partial\vec{x}_0$.  Equation (\ref{B exp}) has a long history, which was reviewed by Stern \cite{Stern:1966} in 1966, its importance was recognized in the 2017 review of magnetic reconnection  by Zweibel and Yamada \cite{Zweibel:review}, and a derivation was given in \cite{Boozer:prevalence}.

A related proof of Equation (\ref{B exp}) is given here for completeness.  The equation $(\partial\vec{B}/\partial t)_{\vec{x}} =\vec{\nabla}\times(\vec{u}\times\vec{B})$ implies
$(\partial \vec{B}/\partial t)_{\vec{x}}=-\vec{B}\vec{\nabla}\cdot\vec{u} + \vec{B}\cdot\vec{\nabla} \vec{u} -\vec{u}\cdot\vec{\nabla}\vec{B}$.  Equation (\ref{dJ/dt}), which is  $(\partial J/\partial t)_{\vec{x}_0}=J\vec{\nabla}\cdot\vec{u} $, and the equation $(\partial \vec{B}/\partial t)_{\vec{x}_0} =(\partial \vec{B}/\partial t)_{\vec{x}}+\vec{u}\cdot\vec{\nabla}\vec{B}$, imply $(\partial (J\vec{B})/\partial t)_{\vec{x}_0}=J\vec{B}\cdot\vec{\nabla} \vec{u}$.  Equation (\ref{B exp}) is valid if 
\begin{eqnarray}
\vec{B}\cdot\vec{\nabla} \vec{u}&=& \frac{\partial }{\partial t} \frac{\partial \vec{x}}{\partial\vec{x}_0}\cdot\vec{B}_0(\vec{x}_0)\\
&=&\left(\frac{\partial \vec{x}}{\partial\vec{x}_0}\cdot\vec{B}_0\right)\cdot\vec{\nabla}\vec{u}
\end{eqnarray}
which can be shown to hold using Equation (\ref{first d tensor(J)/dt}).

When the magnetic field evolves ideally, Equation (\ref{ideal-ev}), the magnetic field $\vec{B}(\vec{x},t)$ at any time $t$ is given by Equation (\ref{B exp}).  This expression for  $\vec{B}(\vec{x},t)$ depends on the initial magnetic field $\vec{B}_0(\vec{x}_0)$ and the Jacobian matrix $\tensor{J}=\partial\vec{x}/\partial\vec{x}_0$ of Lagrangian coordinates, Equation (\ref{def.Lag.coord}). 

Equation (\ref{B exp}) for $\vec{B}(\vec{x},t)$ and Equation (\ref{Jacobian rep}) imply
\begin{eqnarray}
\vec{B}& =&  \frac{\hat{u}\cdot\vec{B}_0}{\Lambda_m\Lambda_s} \hat{U} +  \frac{\hat{m}\cdot\vec{B}_0}{\Lambda_u\Lambda_s} \hat{M} + \frac{\hat{s}\cdot\vec{B}_0}{\Lambda_u\Lambda_m} \hat{S}.\label{B expansion}
\end{eqnarray}

The expression for $\vec{B}$ of Equation (\ref{B expansion}) can be dotted with itself to obtain the square of the magnetic field strength,
\begin{equation}
B^2 =\left(\frac{\hat{u}\cdot\vec{B}_0}{\Lambda_m\Lambda_s}\right)^2  + \left( \frac{\hat{m}\cdot\vec{B}_0}{\Lambda_u\Lambda_s}\right)^2 + \left(\frac{\hat{s}\cdot\vec{B}_0}{\Lambda_u\Lambda_m} \right) \label{B^2 exp}.
\end{equation}

The three terms in Equation (\ref{B^2 exp}) for $B^2$ have fundamentally different time dependencies as $\Lambda_u$ goes to infinity and $\Lambda_s$ goes to zero exponentially.  The term in $B^2$ proportional to  $(\hat{u}\cdot\vec{B}_0)^2$ goes to infinity, the term proportional to $(\hat{s}\cdot\vec{B}_0)^2$ goes to zero exponentially, while the term proportional to $(\hat{m}\cdot\vec{B}_0)^2$ changes only moderately.  The term proportional to $(\hat{u}\cdot\vec{B}_0)^2$ is important in dynamo theory \cite{Tang-Boozer:2000,Thiffeault-Boozer:2003}, but as a system evolves toward a rapidly reconnecting state, an exponentially large increase in the magnetic field strength is not expected.  During the period in which $\hat{u}(\vec{x}_0,t)$ relaxes to its steady-state value, Section \ref{sec:right unit vec relax}, $\hat{u}$ must rotate to a direction orthogonal to $\vec{B}_0$, which implies $\hat{u}\cdot\vec{B}_0\rightarrow0$.  


\section{Evolution of the current density and the Lorentz force \label{sec:j-f ev.} }


\subsection{The current density}

The expanded form for the magnetic field, Equation (\ref{B expansion}), allows the current density, $\vec{j}=\vec{\nabla}\times\vec{B}/\mu_0$, to be determined using the mathematics of general coordinate systems.  Unfortunately, these mathematical methods are not known by most plasma physicists, but a two-page derivation is given in the appendix to \cite{Boozer:RMP}.  

While applying the method of general coordinates to Lagrangian coordinates, $\vec{x}(\vec{x}_0,t)$, the standard convention of using superscripts will be used to number the coordinates.  As before, the three Lagrangian coordinates will be denoted using Latin superscripts $x_0^i$ and the three coordinates in ordinary Cartesian space will be denoted using Greek superscripts $x^\alpha$.  The Jacobian matrix $\tensor{J}\equiv\partial\vec{x}/\partial\vec{x}_0$ has components $J^\alpha_i$, the metric tensor of Lagrangian coordinates, $\tensor{g}\equiv\tensor{J}^\dag\cdot\tensor{J}$ has components $g_{ij}$, and the metric tensor of ordinary cartesian coordinates is the unit tensor $\tensor{1}$, which has the Kronecker delta function, $\delta_{\alpha \beta}$, as components.  With these conventions, dot products are always sums over one superscript and one subscript of the same type, Latin or Greek.

The curl of the magnetic field, when calculated using Lagrangian coordinates, is
\begin{eqnarray}
\vec{\nabla}\times \vec{B} &=& \frac{1}{J} \sum_{ijk} \epsilon^{ijk}\frac{\partial \mathcal{B}_j}{\partial x_0^i} \frac{\partial \vec{x}}{\partial x_0^k},
\end{eqnarray}
where $\epsilon^{ijk}$ is the fully anti-symmetric tensor: $\epsilon^{123}=1$ as are $\epsilon^{312}=1$ and $\epsilon^{231}=1$, which are even permutations of the indices.  The other three permutations are negative, such as  $\epsilon^{213}=-1$.  When two indices are identical $\epsilon^{ijk}$ is zero.  

The $\mathcal{B}_j$ are coefficients of the covariant representation of the magnetic field, $\vec{B} =\sum_j \mathcal{B}_j \vec{\nabla}x_0^j$.  When $\vec{B}$ is known in covariant form, the curl can be simply calculated. The coefficients $\mathcal{B}_j$  can be obtained using the orthogonality relations,
\begin{equation}
\frac{\partial \vec{x}}{\partial x_0^i}\cdot \vec{\nabla} x_0^j = \delta_i^j,
\end{equation}
which follows from the chain rule.  The expanded form of $\vec{B}$, Equation (\ref{B expansion}) implies
\begin{eqnarray}
\mathcal{B}_j &=&\vec{B}\cdot\frac{\partial \vec{x}}{\partial x_0^j} \\
&=& \frac{\Lambda_u \hat{u}\cdot\vec{B}_0}{\Lambda_m\Lambda_s} \hat{u}_j  +   \frac{\Lambda_m \hat{m}\cdot\vec{B}_0}{\Lambda_u\Lambda_s} \hat{m}_j  \nonumber \\
&& +   \frac{\Lambda_s \hat{s}\cdot\vec{B}_0}{\Lambda_u\Lambda_m} \hat{s}_j    \\
&=& h_u \hat{u}_j + h_m \hat{m}_j +h_s \hat{s}_j, \mbox{   where   } \\
h_m &\equiv& \frac{\Lambda_m \hat{m}\cdot\vec{B}_0}{\Lambda_u\Lambda_s} \mbox{  etc.   }
\end{eqnarray}

Let $\hat{e}$ be any of the three eigenvectors, then the contribution of the $\hat{e}\cdot\vec{B}_0$ component of the magnetic field to the current density can be calculated using the representation of the Jacobian matrix expanded in the the left and right eigenvectors of the singular value  decomposition, Equation (\ref{Jacobian rep}).  The current density is 
\begin{eqnarray}
\vec{j}&=&  \frac{\hat{u}\cdot \vec{\nabla}\times(h_e\hat{e}) }{\mu_0\Lambda_m\Lambda_s} \hat{U} + \frac{\hat{m}\cdot \vec{\nabla}\times(h_e\hat{e}) }{\mu_0\Lambda_u\Lambda_s} \hat{M} \nonumber\\
&& +\frac{\hat{s}\cdot \vec{\nabla}\times(h_e\hat{e}) }{\mu_0\Lambda_u\Lambda_m} \hat{S}. \label{current density exp}
\end{eqnarray}


\subsection{The Lorentz force}

The force exerted on the plasma, the Lorentz force, can be calculated by crossing the expression for the current density, Equation (\ref{current density exp}), with Equation (\ref{B expansion}) for the magnetic field.  When the magnetic field strength does not increase exponentially with time, only the $\hat{m}\cdot\vec{B}_0$ term remains non-zero, and the Lorentz force is
\begin{eqnarray}
\vec{f}_L &\equiv& \vec{j}\times\vec{B}\\
&=& \frac{\hat{u}\cdot \vec{\nabla}_0\times(h_m\hat{m}) }{\mu_0\Lambda_m\Lambda_s} \frac{\hat{m}\cdot\vec{B}_0}{\Lambda_u\Lambda_s} \hat{S} \nonumber \\
&& - \frac{\hat{s}\cdot \vec{\nabla}_0\times(h_m\hat{m}) }{\mu_0\Lambda_u\Lambda_m} \frac{\hat{m}\cdot\vec{B}_0}{\Lambda_u\Lambda_s} \hat{U},
\end{eqnarray}
where $\hat{U}\times\hat{M}=\hat{S}$ and $\hat{S}\times\hat{M}=-\hat{U}$.  The term
\begin{eqnarray}
&& \hat{u}\cdot \vec{\nabla}\times(h_m\hat{m}) = \frac{\Lambda_m \hat{m}\cdot\vec{B}_0}{\Lambda_u\Lambda_s} \hat{u}\cdot \vec{\nabla}_0\times\hat{m} \nonumber\\
&& \hspace{1.1in} - \hat{s}\cdot\vec{\nabla}_0\left(\frac{\Lambda_m \hat{m}\cdot\vec{B}_0}{\Lambda_u\Lambda_s}\right),\hspace{0.3in}
\end{eqnarray}
since $\hat{u}\times\hat{m}=\hat{s}$.  The definition of $\hat{u}\cdot (\vec{\nabla}_0\times\hat{m})$ is
\begin{equation}
\hat{u}\cdot (\vec{\nabla}_0\times\hat{m})\equiv  \sum_{ijk} \epsilon^{ijk} \hat{u}_i \frac{\partial \hat{m}_k}{\partial x_o^j}.
\end{equation}
The term
\begin{eqnarray}
&& \hat{s}\cdot \vec{\nabla}_0\times(h_m\hat{m}) = \frac{\Lambda_m \hat{m}\cdot\vec{B}_0}{\Lambda_u\Lambda_s} \hat{s}\cdot \vec{\nabla}_0\times\hat{m} \nonumber\\
&& \hspace{1.1in} + \hat{u}\cdot\vec{\nabla}_0\left(\frac{\Lambda_m \hat{m}\cdot\vec{B}_0}{\Lambda_u\Lambda_s}\right),\hspace{0.3in}.
\end{eqnarray}

The largest term in the Lorentz force is
\begin{eqnarray}
(\vec{f}_L)_{largest} & =&\frac{(\hat{m}\cdot\vec{B}_0 ) \hat{u}\cdot \vec{\nabla}_0\times(h_m\hat{m}) }{\mu_0 \Lambda_u\Lambda_m\Lambda_s} \frac{\hat{S}}{\Lambda_s}  , 
\end{eqnarray}
which predicts an exponentially large force as $t\rightarrow\infty$ unless $\hat{u}\cdot\vec{\nabla}_0 \times (h_m\hat{m})$ relaxes exponentially rapidly to zero.


\subsection{The parallel current density}

The parallel current is given by
\begin{eqnarray}
\mu_0\vec{j}\cdot\vec{B} &=& \left(\frac{\Lambda_m}{\Lambda_u\Lambda_s} \hat{m}\cdot\vec{B}_0\right)^2\hat{m}\cdot (\vec{\nabla}_0\times\hat{m}); \\
\frac{\mu_0\vec{j}\cdot\vec{B}}{B^2}&=&\Lambda_m^2  \hat{m}\cdot (\vec{\nabla}_0\times\hat{m}).  \label{j_||/B}
\end{eqnarray}
As shown in Section \ref{sec:right unit vec relax}, the quantity $\hat{m}\cdot (\vec{\nabla}_0\times\hat{m})$ quickly becomes time independent unless $|\vec{\nabla}\vec{u}|$ scales as $\Lambda_u/\Lambda_m$ or $\Lambda_m/\Lambda_s$.   The middle singular value $\Lambda_m$ changes slowly with respect to time in comparison to the exponential dependences of $\Lambda_u$ and $1/\Lambda_s$.  As shown in Section \ref{sec:field line sep}, the separation between neighboring magnetic field lines is proportional to the variation in $\Lambda_u(\vec{x}_0,t)$ along a given magnetic field line, $d\vec{x}_0/d\ell=\hat{m}(\vec{x}_0)$ at a fixed time $t$.


\subsection{The gradient of $j_{||}/B$}

Equation (\ref{j_||/B}) gives an expression for $j_{||}/B$ as a function of $\vec{x}_0$ at fixed time.  Before calculating the gradient of $j_{||}/B$, the gradient of an arbitrary function $f(\vec{x}_0)$ will be obtained;
 \begin{eqnarray}
 \vec{\nabla}f&=&  \left( \frac{\partial \vec{x}_0}{\partial\vec{x}}\right)^\dag\cdot\frac{\partial f}{\partial \vec{x}_0}, \mbox{   where  }\\
\left( \frac{\partial \vec{x}_0}{\partial\vec{x}}\right)^\dag &=& \frac{\hat{U}\hat{u}}{\Lambda_u}  + \frac{\hat{M}\hat{m}}{\Lambda_m}+ \frac{\hat{S}\hat{s}}{\Lambda_s}, \mbox{    so   } \label{Jacobian inv} \\
\vec{\nabla}f&=& \hat{U} \frac{\hat{u}\cdot\vec{\nabla}_0 f}{\Lambda_u}+ \hat{M} \frac{\hat{m}\cdot\vec{\nabla}_0 f}{\Lambda_m} + \hat{S} \frac{\hat{s}\cdot\vec{\nabla}_0 f}{\Lambda_s},\hspace{0.3in} \label{grad_0 f}
 \end{eqnarray}
 where $\vec{\nabla}_0 f \equiv\partial f/\partial \vec{x}_0$.

The implication is that the gradient of $j_{||}/B$ is exponentially large  in the $\hat{S}$ direction and exponentially larger than the gradients in the other two directions.  Regions of enhanced $j_{||}/B$ are very narrow in the $\hat{S}$ direction.  The gradient in the $\hat{U}$ direction is exponentially small, so regions of enhanced $j_{||}/B$ are very extended in that direction.  Both the $\hat{S}$ and the $\hat{U}$ directions are orthogonal to the magnetic field, which is in the $\hat{M}$ direction. 


\section{Magnetic field lines \label{sec:B-lines} }

Appendix \ref{sec:B corr} is a direct calculation of the change in the magnetic field produced by non-ideal effects when this change is small.  What is found is that this change is to lowest order a change in the direction of the magnetic field,  $\hat{b}\approx\hat{b}_I+\vec{b}_{ni}$, where $\hat{b}_I$ is the direction of the magnetic field would have had if the evolution had been ideal.  Equation (\ref{corr. to b-hat}) for $\vec{b}_{ni}$ shows the change in direction is  proportional to non-ideal part of the electric field times the exponentially large coefficient $\Lambda_u$.

This section shows that an ideal evolution generally leads to an exponentially increasing separation of neighboring field lines with distance along the lines.   This exponentiation in the separation leads to exponential sensitivity to non-ideal effects.

The separation of neighboring magnetic field lines can be determined using the Clebsch potentials $\alpha$ and $\beta$, Section \ref{sec:Clebsch rep}, which have constant values along a magnetic field line, $\vec{B}\cdot\vec{\nabla}\alpha = B \partial\alpha/\partial \ell=0$ and $\vec{B}\cdot\vec{\nabla}\beta = B \partial\beta/\partial \ell=0$.  The Clebsch potentials  can be determined from the starting points of magnetic field lines calculated at a fixed time.  The starting points should be on a surface that is nowhere tangential to $\vec{B}$.  

As discussed in Section \ref{sec:Clebsch rep}, the  equation for the ideal evolution of a magnetic field is satisfied when the Clebsch potentials are functions of the Lagrangian coordinates $\vec{x}_0$, $\alpha(\vec{x}_0)$ and $\beta(\vec{x}_0)$.

\subsection{The gradients of the Clebsch potentials}

The gradients of the Clebsch potentials can be calculated using Equation (\ref{grad_0 f}) for the gradient in ordinary space of a function known in Lagrangian coordinates.
\begin{eqnarray}
\vec{\nabla}\alpha &=& (\hat{u}\cdot\vec{\nabla}_0\alpha)\frac{\hat{U}}{\Lambda_u} + (\hat{m}\cdot\vec{\nabla}_0\alpha)\frac{\hat{M}}{\Lambda_m}  \nonumber \\
&&+ (\hat{s}\cdot\vec{\nabla}_0\alpha)\frac{\hat{S}}{\Lambda_s},  \mbox{    and   } \\
\vec{\nabla}\beta &=& (\hat{u}\cdot\vec{\nabla}_0\beta)\frac{\hat{U}}{\Lambda_u} + (\hat{m}\cdot\vec{\nabla}_0\beta)\frac{\hat{M}}{\Lambda_m}  \nonumber \\
&&+ (\hat{s}\cdot\vec{\nabla}_0\beta)\frac{\hat{S}}{\Lambda_s} 
\end{eqnarray}

The initial magnetic field is $\vec{B}_0 =\vec{\nabla}_0\alpha\times\vec{\nabla}_0\beta$.  Since
\begin{equation}
\vec{\nabla}_0\alpha = \hat{u} (\hat{u}\cdot\vec{\nabla}_0\alpha) + \hat{m} (\hat{m}\cdot\vec{\nabla}_0\alpha) + \hat{s} (\hat{s}\cdot\vec{\nabla}_0\alpha), \end{equation}
\begin{eqnarray}
\hat{u}\cdot\vec{B}_0 &=& (\hat{m}\cdot\vec{\nabla}_0 \alpha)(\hat{s}\cdot\vec{\nabla}_0\beta) \nonumber\\
&&-(\hat{m}\cdot\vec{\nabla}_0\beta)(\hat{s}\cdot\vec{\nabla}_0\alpha);\\
\hat{m}\cdot\vec{B}_0 &=& (\hat{s}\cdot\vec{\nabla}_0\alpha)(\hat{u}\cdot\vec{\nabla}_0\beta)  \nonumber\\
&&-(\hat{s}\cdot\vec{\nabla}_0\beta)(\hat{u}\cdot\vec{\nabla}_0\alpha); \label{m.B0-exp}\\
\hat{s}\cdot\vec{B}_0 &=& (\hat{u}\cdot\vec{\nabla}\alpha)(\hat{m}\cdot\vec{\nabla}_0\beta)\nonumber\\
&&-(\hat{u}\cdot\vec{\nabla}_0\beta)(\hat{m}\cdot\vec{\nabla}_0\alpha), 
\end{eqnarray}
which implies $\vec{B}=\vec{\nabla}\alpha\times\vec{\nabla}\beta$ reproduces Equation (\ref{B^2 exp}) for the expanded form for $\vec{B}$.


 \subsection{Distance between magnetic field lines \label{sec:field line sep} }
 
 Integrations along the field lines of the initial magnetic field, $\vec{B}_0(\vec{x}_0) =\vec{\nabla}_0\alpha \times \vec{\nabla}_0\beta$, can be used to determine the functions $\alpha(\vec{x}_0)$ and $\beta(\vec{x}_0)$.   
 
 Equation (\ref{grad_0 f}) implies the spatial derivatives in ordinary space of $\alpha$ are 
 \begin{eqnarray}
 \hat{U}\cdot\vec{\nabla}\alpha &=&\frac{\hat{u}\cdot \vec{\nabla}_0\alpha}{\Lambda_u} \mbox{  and  } \\
 \hat{S}\cdot\vec{\nabla}\alpha &=&\frac{\hat{s}\cdot \vec{\nabla}_0\alpha}{\Lambda_s}.
 \end{eqnarray}
 When the magnetic field is in the direction $\hat{M}$, the derivative  $\hat{M}\cdot\vec{\nabla}\alpha=0$.  Analogous equations hold for the derivatives of $\beta$.
 
 The implication of these expressions is that $\alpha$ changes weakly in the $\hat{U}$ direction, by an amount proportional to $1/\Lambda_u$.  Magnetic field lines that are infinitesimally separated in $\alpha$ have a separation in the $\vec{U}$ direction proportional to $\Lambda_u(\vec{x}_0,t)$.   
 
 The dependence of the separation along a magnetic field line depends on the variation in  $\Lambda_u$ along the line at fixed time $t$.  This variation is given by Equation (\ref{Lamda_u dependence}) and is comparable to $\Lambda_u$ itself.

 An analogous argument implies the separation of magnetic field lines in the $\hat{S}$ direction is small, proportional to $\Lambda_s(\vec{x}_0,t)$.


\section{Lagrangian coordinates in other fields \label{sec:Lag.appl.} }

\subsection{Mixing in fluids \label{sec:fluid.mix} }

A  closely related situation to fast magnetic reconnection is mixing in stirred fluids---the advection-diffusion problem---in which it is found that the time required for complete mixing depends only logarithmically on the diffusion coefficient.  Aref et al \cite{Aref:2017} have written an informative review---particularly the first section.   The advection-diffusion equation in Cartesian coordinates, $(\partial n/\partial t)_{\vec{x}}+\vec{u}\cdot\vec{\nabla}n=D\nabla^2 n$, when written in Lagrangian coordinates becomes
\begin{equation}
\left(\frac{\partial n}{\partial t}\right)_{\vec{x}_0}=\frac{D}{J} \frac{\partial}{\partial \vec{x}_0} \cdot \left\{ J \left(\frac{\partial\vec{x}_0}{\partial \vec{x}}\right)^\dag\cdot\left(\frac{\partial\vec{x}_0}{\partial \vec{x}}\right) \cdot \frac{\partial n}{\partial \vec{x}_0} \right\}. \label{ad.diff.eq}
\end{equation}
$J$ is the coordinate Jacobian, which is the determinant of the Jacobian matrix $\partial\vec{x}/\partial\vec{x}_0$, and $\partial\vec{x}_0/\partial\vec{x}$ is the matrix inverse of the Jacobian matrix.  As has been shown, the Jacobian matrix and its inverse generally have elements that become exponentially large as time advances, $\propto \exp(\lambda t)$, where $\lambda$ is called a Lyapunov exponent.  In a stirred fluid, the effective diffusion coefficient increases exponentially in magnitude until it becomes sufficiently large to flatten spatial variations in $n$.  As anyone who has observantly stirred coffee or paint is aware, a certain time is required before the mixing rather suddenly occurs.  

Methods of enhancing the effectiveness of stirring in fluids are of practical importance and have received both mathematical and experimental attention.  For example, Boyland, Aref and Stremler \cite{Boyland:2000} have used topological concepts to study various protocols for fluid stirring.

The parallel current $j_{||}$ in a reduced-MHD model \cite{Boozer:prevalence} of an evolving magnetic field, obeys an equation mathematically identical to Equation (\ref{ad.diff.eq}) with $D$ replaced by $\eta/\mu_0$.  The time delay, which is intrinsic to solutions of the advection-diffusion equation, explains the trigger for reconnection.  Both the trigger and the speed of fast magnetic reconnection are easily explained in a  three-dimensional evolving magnetic field.

\subsection{Lagrangian coherent structures}

The enhancement of mixing is undoubtedly the property described by Lagrangian coordinates with which we are most familiar from daily life.  Nevertheless, the Lagrangian description is used not only to explain exponentially enhanced mixing but also to describe spatial regions in moving fluids that form barriers to enhanced mixing.  

Research on describing barriers to enhanced mixing in fluids provides important insights into incomplete magnetic reconnection when the reconnection proceeds at an Alfv\'enic rate.  A review of the situation in fluids has been written by Haller \cite{Haller:2015}, who notes ``\emph{Lagrangian fluid motion is inherently unstable owing to its sensitivity with respect to initial conditions.}"  Although the basic result of this instability is rapid mixing, coherent structures are observed that ``\emph{describe the most repelling, attracting, and shearing material surfaces that form the skeletons of Lagrangian particle dynamics. Uncovering such surfaces from experimental and numerical flow data promises a simplified understanding of the overall flow geometry, an exact quantification of material transport, and a powerful opportunity to forecast, or even influence, large-scale flow features and mixing events.}"  These are called Lagrangian coherent structures.

Lagrangian coherent structures are given by the ridges and trenches of $\Lambda_u(\vec{x}_0,t)$ over some time period.  There is a large literature on the behavior of $\Lambda_u(\vec{x}_0,t)$.  The part of the literature that is associated with magnetic field problems was discussed in Section \ref{sec:right unit vec relax}.   Work in the fluid mechanics community is discussed by Haller \cite{Haller:2015}.


\subsection{The standard map}

A direct simulation of fast magnetic reconnection or of solutions to the advection-diffusion equation will always be beyond the capability of any computer as the non-ideal effects become extremely small, Section 3.8 of \cite{Boozer:prevalence}.  Nevertheless, a laptop computer can be used to study the detailed mathematical properties of Lagrangian coordinates and magnetic field lines by iterating the  standard map \cite{Chirikov:1979}.  The standard map in the $n^{th}$ iteration is
\begin{eqnarray}
\theta_{n+1}=\theta_n + \psi_n; \\
\psi_{n+1}=\psi_n+k \sin\theta_{n+1}; \\
\varphi_{n+1}=\varphi_n +\delta\varphi,
\end{eqnarray}
where $\delta\varphi$ is an arbitrary constant.  A physical interpretation of the variables of the standard map is that $\theta$ is a poloidal angle and $\varphi$ is a toroidal angle with $\psi$ is proportional to the magnetic flux enclosed by a $\psi$ surface.  The rotational transform $\iota=\psi  / \delta\varphi$.  The parameter $k$ is the strength of a perturbation.  Since 
\begin{eqnarray}
\frac{\partial (\psi_{n+1}, \theta_{n+1})}{\partial (\psi_n, \theta_n)}&\equiv& \left(\frac{\partial \psi_{n+1}}{\partial \psi_n}\right)_{\theta_n}\left(\frac{\partial \theta_{n+1}}{\partial \theta_n}\right)_{\psi_n} \nonumber\\
&& -  \left(\frac{\partial \theta_{n+1}}{\partial \psi_n}\right)_{\theta_n}\left(\frac{\partial \psi_{n+1}}{\partial \theta_n}\right)_{\psi_n} \nonumber\\
&=&1,
\end{eqnarray}
this map represents the divergence-free character of the magnetic field and can be iterated to plot possible trajectories of magnetic field lines---or of a divergence-free flow.  It has the oddity that it is not only periodic $\theta$ but also in $\psi$.  That is $\psi+2\pi$ obeys the same equation as $\psi$.  For simplicity of discussion, $k\geq0$ is assumed.

John Greene \cite{Greene:1979} found that for  $k<0.971635\cdots$ the field lines cover only limited range of $\psi$.  That is, $\psi$ remains within the $2\pi$ periodicity of its initial value.  For larger $k$, some lines, though not all, would cover an unbounded range of $\psi$.  Even for $k<<1$, small regions exist in $(\psi,\theta)$ space in which neighboring magnetic field lines separate exponentially as the map is iterated.  But, a barrier with a complicated shape exists, crudely at $\psi=\pi$, which field lines cannot cross for $k<0.971635\cdots$.  For $k$ slightly greater than this critical value, the remnants of the barrier exist in the form of a cantorus  \cite{Meiss:1984} or a Lagrangian coherent structure, which greatly slows the mixing of trajectories from its two sides.  Even when $k=2$, substantial but isolated regions exist in $(\psi,\theta)$ space in which field lines do not exponentially separate and enhanced mixing would not occur.  Indeed, each of these bounded regions, which are called islands, is surrounded by its own cantorus.  All of these complicated structures have analogues in the behavior of magnetic field lines or of divergence-free flows.


\section{Discussion \label{sec:Discussion} }

Much of the existing literature on magnetic reconnection is two dimensional, which in three-dimensional space implies a continuous perfect symmetry in the third direction.  Two-dimensional models are inadequate when non-ideal effects are weak:

\begin{itemize}

\item  Where is the locus of reconnection?

In two-dimensional theories, reconnection occurs where the two-dimensional part of the magnetic field vanishes independent of the magnitude of the field in the direction of symmetry.   In a true three-dimensional system, such as a tokamak subjected to weak three dimensional perturbations, the reconnection occurs at rational surfaces on which magnetic field lines close on themselves.  On irrational magnetic surfaces or in stochastic regions, the response to perturbations is fundamentally different because the perturbation is spread over the full volume covered by a magnetic field line---or more precisely the volume covered by an Alfv\'en wave moving along a field line during the time taken to produce the perturbation.  The third coordinate is ignorable only when the symmetry of the original magnetic field and the perturbation holds over the spatial scale sampled by an Alfv\'en wave during the time required to turn on the perturbation.

\item  Are exponentially-large terms ignorable? 

For all but exceptional magnetic evolutions, non-ideal effects, which are represented by $\mathcal{E}_{ni}$, are multiplied by an exponentially large factor, $\Lambda_u=e^{\lambda_u t}$, to determine the deviation $\vec{b}_{ni}$ in the magnetic field direction from the direction it would have had in an ideal evolution.  Evolutions that preserve a continuous perfect symmetry are examples of the exceptional evolutions for which the Lyapunov exponent $\lambda_u=0$.

\end{itemize}

Three-dimensionality is irrelevant when non-ideal effects are sufficiently strong that reconnection occurs before Alfv\'en waves can propagate a significant distance or before $\Lambda_u$ can become large.  But, these conditions are very restrictive on the applicability of two-dimensional reconnections theories.

The obvious argument for the use of two-dimensional theories is mathematical simplicity.  But in conversations, those who publish work based on two-dimensional analysis claim agreement with observations shows correctness.  Nevertheless, no doubt is expressed on the validity of Maxwell's equations or of mathematics.  The implication is that it should be mathematically possible to show a more general applicability of two-dimensional models to three dimensional space.

The Lagrangian solutions to the magnetic evolution equations that were developed in this paper give important constraints on the validity of two-coordinate models--the weaker the non-ideal effects the more restrictive are the constraints.   The smallness of $\mathcal{E}_{ni}$, the non-ideal part of the electric field, when fast magnetic reconnection occurs makes the conservation of magnetic helicity obvious but the transfer of energy from the magnetic field to the plasma subtle.  These issues are discussed in \cite{Boozer:part.acc.}.  

Numerical simulation is a way forward for understanding magnetic reconnection in three-dimensional space.  But, no computer will ever be able to carry out a direct numerical simulation  of fast magnetic reconnection in the limit as non-ideal effects approach zero.  See Section 3.8 of \cite{Boozer:prevalence}.  Many reconnection problems of importance will remain beyond the power of direct computations.   What can be done is to couple numerical results that are sufficiently restricted to be consistent with existing computational power to extrapolate results to relevant regimes using constraints such as those obtained through methods based on Lagrangian coordinates.

What is clear is that computational and theoretical work on magnetic reconnection must move beyond the two-dimensional models that have dominated the field for more than sixty years and still form the basis of most papers being published.


\appendix



\section{Near ideal correction to $\vec{B}$ \label{sec:B corr}}

When the electric field has the non-ideal form $\vec{E}+\vec{u}\times\vec{B} = -\vec{\nabla}\Phi + \mathcal{E}_{ni}(\alpha,\beta,t)\vec{\nabla}\ell$, the equations derived in Section \ref{sec:Clebsch rep} imply Clebsch potentials have the form
\begin{eqnarray}
\alpha &=& \alpha_I(\vec{x}_0) - \frac{\partial \mathcal{A}_{ni}}{\partial \beta} \ell \label{alpha ev}\\
\beta &=& \beta_I(\vec{x}_0) + \frac{\partial \mathcal{A}_{ni}}{\partial \alpha} \ell  \label{beta ev}\\
\mathcal{A}_{ni} &\equiv& - \int_0^t \mathcal{E}_{ni}(\alpha,\beta,t)dt, \label{A_ni}
\end{eqnarray}
where $\mathcal{A}_{ni}$ is the non-ideal part of the vector potential, which will be assumed to produce only a small change to the magnetic field from its ideal form $\vec{B}_I(\vec{x},t)=\vec{\nabla}\alpha_I \times \vec{\nabla}\beta_I$.


\subsection{Ideal Clebsch potentials as Lagrangian coordinates}

The perturbative calculation of the deviation of the magnetic field from that of an ideal evolution uses the Clebsch potentials of an ideal evolution $\alpha_I$ and $\beta_I$ as Lagrangian coordinates.  That this is possible is implied by Section \ref{sec:Clebsch rep}.   A third Lagrangian coordinate is required.  What will be shown is that $\ell$, the distance along field lines, can be used.  To show this, the fundamental definition of Clebsch potentials for a given ideally-evolving magnetic field $\vec{B}_I(\vec{x},t)$ must be considered.  This definition uses a two-dimensional Clebsch surface in three-dimensional space.

Let $\vec{x}_{c}(\rho,\beta_I,t)$ be a Clebsch surface in three space, which means that it is nowhere tangential to the magnetic field in the region of interest.  The two coordinates that give positions on the Clebsch surface are $\rho$ and $\beta_I$.   Ordinary Cartesian coordinates $\vec{x}$ are functions of where a particular magnetic field line penetrates the Clebsch surface and the distance $\ell$ that the field line must be followed to reach that surface.  At each instant of time, the Cartesian coordinates can be given as $\vec{x}(\rho,\beta_I,\ell,t)$ with $\partial \vec{x}/\partial\ell = \hat{b}_I$ with $\hat{b}_I$  the unit vector along the ideal magnetic field.  Since $\rho$ and $\beta$ are constant along the magnetic field, $\vec{B}_I=f\vec{\nabla}\rho\times\vec{\nabla}\beta_I$.  The constraint $\vec{\nabla}\cdot\vec{B}_I=0$ requires $f$ satisfy $\partial f/\partial \ell=0$.  The implication is that $\vec{B}_I=\vec{\nabla}\alpha_I\times\vec{\nabla}\beta_I$, where $\partial \alpha_I(\rho,\beta_I,t)/\partial\rho =f(\rho,\beta_I,t)$.

Although $\alpha_I$ and $\beta_I$ can be taken to be constant in Lagrangian coordinates, it remains to be shown that $\ell$ can be taken to be a Lagrangian coordinate.  As with any function of position and time, the time derivative of $\ell$ in Lagrangian coordinates is $(\partial \ell/\partial t)_{\vec{x}_0}= (\partial \ell/\partial t)_{\vec{x}}+\vec{u}\cdot\vec{\nabla}\ell$.  Since $\ell$ is to be a Lagrangian coordinate, $(\partial \ell/\partial t)_{\vec{x}_0}=0$, and the Clebsch surface must be given the velocity $\vec{v}_{c}$ required for $(\partial \ell/\partial t)_{\vec{x}}=-\vec{u}_\bot\cdot\vec{\nabla}\ell$ with $\ell=0$ on the Clebsch surface.  In order not to change $\alpha_I$ and $\beta_I$ while moving the Clebsch surface, the motion must be along the magnetic field, so the required velocity of the Clebsch surface is
\begin{equation}
\vec{v}_{c} =\big( \hat{b}_I \vec{u}\cdot\vec{\nabla}\ell\big)_{c}.
\end{equation}
The quantities that define the velocity of the Clebsch surface are evaluated on the instantaneous Clebsch surface. 


\subsection{Calculation of the non-ideal field}

Using $(\alpha_I,\beta_I,\ell)$ as Lagrangian coordinates, Equations (\ref{alpha ev}) to (\ref{A_ni}) imply the magnetic field is
\begin{eqnarray}
\vec{B} &=& \vec{B}_I - \frac{\partial \mathcal{A}_{ni}}{\partial \beta_I} \vec{\nabla}\ell \times \vec{\nabla}\beta_I + \frac{\partial \mathcal{A}_{ni}}{\partial \alpha_I} \vec{\nabla}\alpha_I  \times \vec{\nabla}\ell \nonumber \\
&=& \vec{B}_I + B_I \left( \frac{\partial \mathcal{A}_{ni}}{\partial \beta_I} \frac{\partial \vec{x}}{\partial\alpha_I} -  \frac{\partial \mathcal{A}_{ni}}{\partial \alpha_I} \frac{\partial \vec{x}}{\partial\beta_I} \right),
\end{eqnarray}
using the dual relations, $J_c(\vec{\nabla}\ell\times\vec{\nabla}\alpha_I)=\partial\vec{x}/\partial\beta_I$ with $1/J_c=(\vec{\nabla}\alpha_I\times\vec{\nabla}\beta_I)\cdot\vec{\nabla}\ell=B_I$ since $\vec{B}_I\cdot(\partial\vec{x}/\partial\ell)=B_I$.   The dual relations are derived and explained in the Appendix to \cite{Boozer:RMP}.

The freedom of canonical transformations contained in $g_a(\alpha,\beta,t)$ implies the two ideal Clebsch potentials can be chosen as
\begin{eqnarray}
\vec{\nabla}\alpha_I&=& \frac{\hat{s}\cdot\vec{\nabla}_0\alpha_I}{\Lambda_s}\hat{S};  \label{grad alpha}\\
\vec{\nabla}\beta_I&=& \frac{\hat{u}\cdot\vec{\nabla}_0\beta_I}{\Lambda_u}\hat{U}; \label{grad beta} \\
\vec{B}_I &=& \frac{(\hat{s}\cdot\vec{\nabla}_0\alpha_I)(\hat{u}\cdot\vec{\nabla}_0\beta_I)}{\Lambda_u\Lambda_s} \hat{M}.
\end{eqnarray}
Consistency with Equation (\ref{B expansion}) implies
\begin{equation}
(\hat{s}\cdot\vec{\nabla}_0\alpha_I)(\hat{u}\cdot\vec{\nabla}_0\beta_I)=B_0,
\end{equation}
the initial magnetic field strength.

The orthogonality relations, which are derived in  the Appendix to \cite{Boozer:RMP}, can then be shown to imply
\begin{eqnarray}
\frac{\partial \vec{x}}{\partial\alpha_I}&=& \frac{\Lambda_s \hat{S}}{\hat{s}\cdot\vec{\nabla}_0\alpha_I} + \frac{B_\alpha}{B_I} \hat{M} \label{x alpha}\\
\frac{\partial \vec{x}}{\partial\beta_I}&=& \frac{\Lambda_u \hat{U}}{\hat{u}\cdot\vec{\nabla}_0\beta_I} +\frac{B_\beta}{B_I} \hat{M}.
\end{eqnarray}
$B_\alpha$ and $B_\beta$ are coefficients in the expansion of  the ideal magnetic field in the gradients of the Clebsch coordinates,
\begin{eqnarray}
\vec{B}_I& =&B_I\vec{\nabla}\ell+B_\alpha\vec{\nabla}\alpha_I +B_\beta\vec{\nabla}\beta  \nonumber\\
&=& B_I\hat{M}. \label{cov.Clebsch}
\end{eqnarray}
Equation (\ref{x alpha}) follows using Equations (\ref{grad alpha}) and (\ref{grad beta}) by writing
\begin{eqnarray}
&&\frac{\partial\vec{x}}{\partial\alpha_I}=c_u\hat{U} +c_s\hat{S} + c_m \hat{M}\\
&& \frac{\partial\vec{x}}{\partial\alpha_I}\cdot\vec{\nabla}\alpha_I=1  \mbox{   so } c_s =\frac{\Lambda_s}{\hat{s}\cdot \vec{\nabla}_0\alpha_I} \\
&& \frac{\partial\vec{x}}{\partial\alpha_I}\cdot\vec{\nabla}\beta_I=0  \mbox{   so } c_u =0 \\
&& \frac{\partial\vec{x}}{\partial\alpha_I}\cdot\hat{M} = \frac{B_\alpha}{B},
\end{eqnarray}
where the expression for $(\partial\vec{x}/\partial\alpha_I)\cdot\hat{M}$  follows from Equation (\ref{cov.Clebsch}) since $(\partial\vec{x}/\partial\alpha_I)\cdot\vec{\nabla}\ell=0$.


As $\Lambda_u\rightarrow\infty$, the expression for the magnetic field $\vec{B}$ has the form
\begin{eqnarray}
\vec{B}&\rightarrow& \vec{B}_I - B_I\frac{\partial\mathcal{A}_{ni}}{\partial \alpha_I} \frac{\Lambda_u \hat{U}}{\hat{u}\cdot\vec{\nabla}_0\beta_I }; \\
\hat{s}\cdot\vec{\nabla}_0 \mathcal{A}_{ni} &=& (\hat{s}\cdot\vec{\nabla}_0\alpha_0)\frac{\partial \mathcal{A}_{ni}}{\partial\alpha_I},   \mbox{   so   } \\
\vec{B}&\rightarrow& \vec{B}_I -B_I \hat{U} \frac{\Lambda_u }{B_0} \hat{s}\cdot\vec{\nabla}_0 \mathcal{A}_{ni}.
\end{eqnarray} 
Since the non-ideal correction to the magnetic field is an orthogonal direction to the ideal field, the correction to the field strength is negligible in a first order calculation.

In the asymptotic limit as $\Lambda_u\rightarrow\infty$, the deviation in the magnetic field line direction due to non-ideal effects is 
\begin{eqnarray}
\vec{b}_{ni} &=& - \Lambda_u\frac{ \hat{s}\cdot\vec{\nabla}_0 \mathcal{A}_{ni}}{B_0} \hat{U} \label{corr. to b-hat}
\end{eqnarray}

\section{Current density in an ideal-evolution model  \label{RMHD}  }

Consider a model  in $(x,y,z)$ Cartesian coordinates  \cite{Boozer:prevalence}  in which initially straight magnetic field lines go from a perfectly conducting stationary wall at $z=0$ to another perfectly conducting but moving wall at $z=L$.  The plasma between the walls is assumed to be ideal, $\vec{E}+\vec{u}\times\vec{B}=0$.  The flow velocity $\vec{u}_w$ of the $z=L$ wall is assumed to be on a scale $a<<L$ and very slow compared to $(a/L)V_A$, so Alfv\'en waves can keep the magnetic field in a force-free, $\vec{j}\times\vec{B}=0$, state.  The evolution of the magnetic field in the plasma, $0<z<L$, is determined by the vorticity of the slow, divergence-free, flow in the wall $\Omega_w(x,y,t)=\hat{z}\cdot\vec{\nabla}\times\vec{u}_w$.  In this model,  the distance along the field lines $\ell$ equals $z$, 
\begin{eqnarray}
\vec{u}&=& \vec{\nabla}\times (\phi \hat{z}),  \mbox{   and   } \\
\nabla_\bot^2 \phi &=&- \Omega,  \\
\vec{B} &=& B_0\big(\hat{z}+\vec{\nabla}\times(H\hat{z})\big),\\
\nabla_\bot^2H &=& -\mathcal{K},
\end{eqnarray}
where  $\nabla_\bot^2 \equiv \partial^2/\partial x^2 + \partial^2/\partial y^2$ and $H(x,y,z,t)$ gives the magnetic field lines $dx/dz=\partial H/\partial y$ and $dy/dz=-\partial H/\partial x$.  A solution to the equations is
\begin{eqnarray}
\Omega(x,y,z,t) &=& \Omega_w(\alpha,\beta,t)\frac{z}{L}, \label{Omega soln} \\
\left(\frac{\partial \mathcal{K}}{\partial t}\right)_{\alpha,\beta} &=&\frac{\Omega_w(\alpha,\beta,t)}{L}, \label{K soln}
\end{eqnarray}
where the distribution of parallel current, $\mathcal{K}\equiv\mu_0j_{||}/B$, is independent of $\ell$.  When $\phi_w$ is independent of time $\mathcal{K}$ increases linearly with time.    

The magnetic field line labels $\alpha(x,y,z,t)$ and $\beta(x,y,z,t)$ are time independent functions of the $(x,y)$ coordinates in the $z=0$ stationary wall but are generally complicated functions of $(x,y,z,t)$ for $z$ in the range  $0\leq z\leq L$.  On the perfectly conducting wall at $z=L$, the magnetic field line labels are equivalent to two-dimensional Lagrangian coordinates for the flow of the wall. $\partial \alpha(x,y,L,t)/\partial t = - \vec{u}_w\cdot\vec{\nabla}\alpha$ and $\partial \beta(x,y,L,t)/\partial t = - \vec{u}_w\cdot\vec{\nabla}\beta$.

The solution given by Equations (\ref{Omega soln}) and (\ref{K soln}) may become unstable; when this occurs the relevant solution is more complicated.

When the flow velocity in the $z=L$ wall is written in terms of its stream function, $\vec{u}_w=\vec{\nabla}\times\big(\phi_w(x,y,t)\hat{z}\big)$, the streamlines of the wall flow are $dx/dt = \partial \phi_w/\partial y$ and $dy/dt=-\partial\phi_w/\partial x$, which are of identical form to the equations of classical mechanics of one and a half degrees of freedom, $H(p,q,t)$.  When $\partial \phi_w/\partial t=0$, the stream lines remain on constant $\phi_w$ surfaces, and the regions in the $z=L$ surface in which the streamlines increase the separation by $e^{\sigma}$ over time occupy an $e^{-2\sigma}$ fraction of the surface area.  When $\phi_w$ is time dependent, then generally a large fraction of the area is occupied by streamlines that increase their separation exponentially in time, as $e^{\lambda_st}$, where $\lambda_s>0$ is called a Lyapunov exponent.  Even simple stream functions can produce complicated spatial distributions of $j_{||}/B$.  An example is a Fourier representation with a periodicity distance $a$ that has at least two terms,
\begin{equation}
\phi_w = \sum_{mn} \phi_{mn} \sin\left(n\frac{x}{a} +m \frac{y}{a} -\omega_{mn}t\right),
\end{equation}
where the $\phi_{mn}$ and the $\omega_{mn}$ are distinct constants.


\section*{Acknowledgements}

This material is based upon work supported by the U.S. Department of Energy, Office of Science, Office of Fusion Energy Sciences under Award Numbers DE-FG02-95ER54333, DE-FG02-03ER54696, DE-SC0018424, and DE-SC0019479.



\begin{thebibliography}{99}

\bibitem{Newcomb} W. A. Newcomb, \emph{Motion of magnetic lines of force}, Ann. Phys. \textbf{3}, 347 (1958).


\bibitem{Parker:2012} E. N. Parker, \emph{Singular magnetic equilibria in the solar x-ray corona}, Plasma Phys. Control. Fusion \textbf{54}, 124028 (2012). 


\bibitem{Rappazzo-Parker} A. F. Rappazzo and E. N. Parker,  \emph{ Current sheets formation in tangled coronal magnetic fields}, Ap. J. Lett. \textbf{773}, L2 (2013).

 \bibitem{Boozer:B-line.sep} A. H. Boozer, \emph{Separation of magnetic field lines}, Phys. Plasmas \textbf{19}, 112901 (2012).
 
 
 \bibitem{Harris} E. G. Harris, \emph{On a plasma sheath separating regions of oppositely directed magnetic fields}, Nuovo Cimento, \textbf{23}, 115-121, (1962).


\bibitem{Boozer:current sheets} A. H. Boozer,  \emph{Formation of current sheets in magnetic reconnection}, Phys. Plasmas \textbf{21}, 072907 (2014).

 
\bibitem{Boozer:prevalence} A. H. Boozer,  \emph{Why fast magnetic reconnection is so prevalent}, J. Plasma Phys. \textbf{84}, 715840102 (2018).

\bibitem{Longcope-Strauss}  D.W. Longcope and H. R. Strauss, \emph{The form of ideal current layers in line-tied magnetic fields}, Ap.J. \textbf{437} 851 (1994).


\bibitem{Boozer:NF3D} A. H. Boozer, \emph{Non-axisymmetric magnetic fields and toroidal plasma confinement}, Nucl. Fusion \textbf{55}, 025001 (2015).


 \bibitem{Heyvaerts-Priest:1983} J. Heyvaerts and E. R. Priest, \emph{Coronal heating by phase mixed shear Alfv\'en waves}, Astron. Astrophys. \textbf{117}, 220 (1983).

\bibitem{Similon:1989} P.L. Similon and R. N. Sudan, \emph{Energy-dissipation of Alfv\'en-wave packets deformed by irregular magnetic-fields in solar-coronal arches},  Ap. J. \textbf{336}, 442-453 (1989).


\bibitem{Cassak:2017} P. A. Cassak, Y.-H. Liu, and M. A. Shay, \emph{A review of the 0.1 reconnection rate problem }, J. Plasma Phys. \textbf{83}, 15830501 (2017).

\bibitem{Zweibel:review} E. G. Zweibel and M. Yamada, \emph{Perspectives on magnetic reconnection}, Proc. R. Soc. A \textbf{472}, 20160479 (2016).


\bibitem{Loureiro:2016} N. F. Loureiro and D. A. Uzdensky, \emph{Magnetic reconnection: from the Sweet-Parker model to stochastic plasmoid chains}, Plasma Phys. and Control. Fusion \textbf{58}, 014021 (2016).


\bibitem{Sweet:1958} P. A. Sweet, \emph{The Neutral Point Theory of Solar Flares} in IAU Symposium 6, Electromagnetic Phenomena in Cosmical Physics, ed. B. Lehnert (Dordrecht: Kluwer, 1958) p. 123.

\bibitem{Parker:1957} E. N. Parker, \emph{Sweet's mechanism for merging magnetic fields in conducting fluids}, Journal of Geophysical Research. \textbf{62}, 509 (1957).


\bibitem{Liu:2017} Y.-H. Liu, M. Hesse, F. Guo, W. Daughton, H. Li, P. A. Cassak, and M. A. Shay, \emph{Why does steady-state magnetic reconnection have a maximum local rate of order 0.1?}, Phys. Rev. Lett. \textbf{118}, 085101 (2017).


\bibitem{Boozer-Pomphrey} A. H. Boozer and N. Pomphrey, \emph{Current density and plasma displacement near perturbed rational surfaces}, Phys. Plasmas \textbf{17}, 110707  (2010).

 \bibitem{Hahm-Kulsrud} T. S. Hahm and R. M. Kulsrud, \emph{Forced magnetic reconnection}, Phys. Fluids \textbf{28}, 2412 (1985).
 
\bibitem{Turbulent reconnection} A. Lazarian, G. Eyink, E. Vishniac, and G. Kowal, \emph{Turbulent reconnection and its implications}, Philos. Trans. Royal Soc. A  \textbf{373}, 20140144 (2015).
 
 \bibitem{Low:2015} B. C. Low, \emph{Field topologies in ideal and near-ideal magnetohydrodynamics and vortex dynamics?}, Science China Physics, Mechanics and Astronomy, \textbf{58}, Issue 1: 015201 (2015).
 
\bibitem{Dewar:2017} R. L. Dewar, S. R. Hudson, A. Bhattacharjee, and Z. Yoshida, \emph{Multi-region relaxed magnetohydrodynamics in plasmas with slowly changing boundaries---Resonant response of a plasma slab}, Phys. Plasmas \textbf{24}, 042507 (2017).

\bibitem{de Vries:2016}  P.C. de Vries, G. Pautasso, E. Nardon, P. Cahyna, S. Gerasimov,
J. Havlicek, T.C. Hender, G.T.A. Huijsmans, M. Lehnen, M. Maraschek, T. Markovi\v{c}, J.A. Snipes and the COMPASS Team, the ASDEX Upgrade Team and JET Contributors, \emph{Scaling of the MHD perturbation amplitude required to trigger a disruption and predictions for ITER},  Nucl. Fusion \textbf{56}, 026007 (2016).


\bibitem{Boozer:pivotal} A. H. Boozer, \emph{Pivotal issues on relativistic electrons in ITER}, Nucl. Fusion \textbf{58}, 036006 (2018).


\bibitem{Boozer:2019} A. H. Boozer, \emph{Magnetic surface loss and electron runaway}, Plasma Phys. Control. Fusion \textbf{61}, 024002 (2019). 

 
 \bibitem{Stern:1970} D. P. Stern, \emph{Euler potentials,} Am. J. Phys. \textbf{38}, 494 (1970).

 \bibitem{Goldhirsch:1987} I.Goldhirsch, P.-L. Sulem, and S. A. Orszag,  \emph{Stability and Lyapunov stability of dynamical systems: a differential approach and numerical method}, Physica D \textbf{27}, 311 (1987).


\bibitem{Tang-Boozer:1996} X. Z. Tang and A. H. Boozer, \emph{Finite time Lyapunov exponent and advection-diffusion equation}, Physica D \textbf{95},  283 (1996).


\bibitem{Tang-Boozer:1997} X. Z. Tang, \emph{Hamiltonian structure of Hamiltonian chaos}, Physics Letters A \textbf{236} (1997) 476 (1997).

 
 \bibitem{Tang-Boozer:2000}  X. Z. Tang and A. H. Boozer, \emph{Anisotropies in magnetic field evolution and local Lyapunov exponents}, Phys. Plasmas \textbf{7}, 1113 (2000).


\bibitem{Thiffeault:2002} J.-L. Thiffeault, \emph{Derivatives and constraints in chaotic flows: asymptotic behaviour and a numerical method}, Physica D \textbf{172}, 139 (2002). 

 
\bibitem{Thiffeault-Boozer:2003} J.-L. Thiffeault and A. H. Boozer, \emph{The onset of dissipation in the kinematic dynamo}, Phys. Plasmas \textbf{10}, 259 (2003)


\bibitem{Stern:1966}  D. P. Stern, \emph{The motion of magnetic field lines}, Space Sci. Rev. \textbf{6}, 147 (1966).


\bibitem{Boozer:RMP} A. H. Boozer, \emph{Physics of magnetically confined plasmas}, Rev. Mod. Phys. \textbf{76}, 1071 (2004).


\bibitem{Aref:2017} H. Aref, J. R. Blake, Marko Budisi\'c, S. S. S. Cardoso, J. H. E. Cartwright, H. J. H. Clercx, K. El Omari, U. Feudel, R. Golestanian, E. Gouillart, G. J. F. van Heijst, T. S. Krasnopolskaya, Y. Le Guer, R. S. MacKay, V. V. Meleshko, G. Metcalfe, I. Mezi\'c, A. P. S. de Moura, O. Piro, M. F. M. Speetjens, R. Sturman, J.-L. Thiffeault, and I. Tuval, \emph{Frontiers of chaotic advection},  Rev. Mod. Phys. \textbf{89}, 025007 (2017).

\bibitem{Boyland:2000} P. L. Boyland, H. Aref, and M. A. Stremler, \emph{Topological fluid mechanics of stirring}, J. Fluid Mech. \textbf{403} 277 (2000).

\bibitem{Haller:2015} G. Haller, \emph{Lagrangian Coherent Structures}, Annu. Rev. Fluid Mech. \textbf{47}, 137 (2015).

\bibitem{Chirikov:1979} B. V. Chirikov, \emph{A universal instability of many-dimensional oscillator systems}, Phys. Rep. \textbf{52}, 264 (1979).

\bibitem{Greene:1979} J. M. Greene, \emph{Method for determining a stochastic transition}, J. Math. Phys. \textbf{20}, 1183 (1979).
 

\bibitem{Meiss:1984} R. S. Mackay, J. D. Meiss, and I. C. Percival, \emph{Transport in Hamiltonian Systems},  Physica \textbf{13}D, 55 (1984).

\bibitem{Boozer:part.acc.} A. H. Boozer, \emph{Fast magnetic reconnection and particle acceleration}, arXiv:1902.10860v1, 28 Feb 2019.

\end{thebibliography}
\end{document}